\documentclass[twocolumn]{aastex62}

\usepackage{amsmath}

\usepackage{color}

\begin{document}

\title{Introducing CGOLS: The Cholla Galactic OutfLow Simulation Suite}

\correspondingauthor{Evan Schneider}
\email{es26@astro.princeton.edu}

\author[0000-0001-9735-7484]{Evan E. Schneider\footnotetext{}}
\altaffiliation{Hubble Fellow}
\affiliation{Department of Astrophysical Sciences, Princeton University, 4 Ivy Lane, Princeton, NJ 08544, USA}

\author[0000-0002-4271-0364]{Brant E. Robertson}
\affiliation{Department of Astronomy and Astrophysics, University of California, Santa Cruz, 1156 High Street, Santa Cruz, CA 95064, USA}

\begin{abstract}

We present the Cholla Galactic OutfLow Simulations (CGOLS) suite, a set of extremely high resolution global simulations of isolated disk galaxies designed to clarify the nature of multiphase structure in galactic winds.  Using the GPU-based code Cholla, we achieve unprecedented resolution in these simulations, modeling galaxies over a 20 kpc region at a constant resolution of 5 pc. The simulations include a feedback model designed to test the effects of different mass- and energy-loading factors on galactic outflows over kiloparsec scales. In addition to describing the simulation methodology in detail, we also present the results from an adiabatic simulation that tests the frequently adopted analytic galactic wind model of Chevalier \& Clegg. Our results indicate that the Chevalier \& Clegg model is a good fit to nuclear starburst winds in the nonradiative region of parameter space. Finally, we investigate the role of resolution and convergence in large-scale simulations of multiphase galactic winds. While our largest-scale simulations do show convergence of observable features like soft X-ray emission, our tests demonstrate that simulations of this kind with resolutions greater than $10\,\mathrm{pc}$ are not yet converged, confirming the need for extreme resolution in order to study the structure of winds and their effects on the circumgalactic medium.

\end{abstract}


\section{Introduction} \label{sec:intro}

Galactic outflows are recognized as an important driver of galaxy evolution.  Outflows regulate the mass of galaxies and their baryon to dark matter ratios and enrich the circumgalactic and intergalactic media (CGM and IGM) with metals \citep{Dekel86, Katz96, Springel03, Scannapieco08, Oppenheimer08, Mashchenko08, Dave11, Hopkins14, Vogelsberger14, Schaye15}. The input of mass and energy from massive stars, particularly via supernovae, is thought to be the major driver of galactic outflows in low-mass star-forming galaxies \citep{Heckman90, Hopkins12}. Understanding the process by which stellar feedback drives gas out of galaxies is therefore a fundamental goal of current galaxy evolution theories. 

Numerical simulations have become a critical tool in our theoretical understanding of galaxy evolution, as they can model and predict nonlinear processes that cannot be easily investigated analytically. Supernova-driven galactic winds are certainly such a process, given their complex observed structures and the number of relevant physical processes governing their evolution. Hydrodynamics, radiation, magnetic fields, conduction, and gravity may all play important roles, and the effects of many of these processes have been investigated via 2D simulations \citep[e.g.][]{MacLow89, Strickland00}, 3D simulations of the inner region of starburst galaxies \citep[e.g][]{Cooper08, Sarkar15}, global galaxy simulations with static mesh refinement \citep[e.g.][]{Fielding17b}, local box simulations of dense clouds within a rarefied hot wind \citep[e.g][]{Schiano95, Marcolini05, Melioli05, Cooper09, McCourt15, Scannapieco15, Bruggen16, Schneider17}, and simulations of patches of a galaxy's interstellar medium (ISM) and regions above the disk \citep[e.g.][]{Joung06, Kim13, Walch15, Martizzi16, Li17}.

Simulating starburst-driven winds has its own challenges, however. In addition to the computational expense of including the relevant physics, the range of spatial scales involved is daunting. Critical parameters of winds include their mass and energy loading, $\dot{M} = \dot{M}_\mathrm{wind}/\dot{M}_\mathrm{SFR}$ and $\dot{E} = \dot{E}\mathrm{wind} / \dot{E}_\mathrm{SN}$, which are set on the scale of individual supernova bubbles on the order of a few parsecs \citep{Kim14, Martizzi14, Kim17}. The global distribution of sources throughout the disk may have a large effect on the subsequent evolution of the outflows and requires global galaxy simulations to capture \citep{Martizzi16, Vijayan17}. Following the winds out into the CGM makes the problem even more challenging, as the scales now approach the virial radius of a galaxy ($\sim 100\,\mathrm{kpc}$), while individual cool clouds may still exist with size scales $<< 1\,\mathrm{kpc}$ \citep{Fielding17a, McCourt18}.

\begin{figure}
\includegraphics[width=\linewidth]{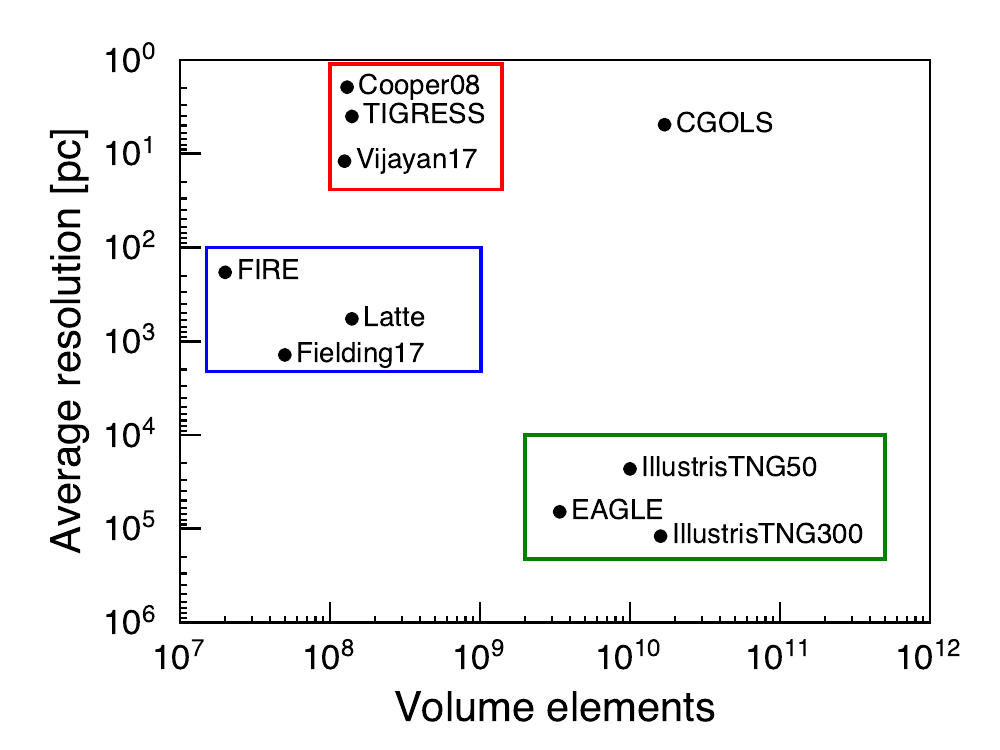}
\caption{Comparison of the average resolution for a representative group of simulations used to study galactic outflows. Average resolution is calculated as the total box size divided by the total number of hydrodynamic volume elements (particles or cells). Different types of simulations are outlined in colored boxes: red shows isolated patch and global galaxy simulations, blue shows zoom and static mesh refinement simulations, and green shows cosmological simulations. Simulation data are from \cite{Cooper08}, \citet[FIRE]{Hopkins14}, \citet[EAGLE]{Schaye15}, \citet[Latte]{Wetzel16}, \cite{Fielding17b}, \cite{Vijayan17}, \citet[Illustris TNG]{Nelson17}, \citet[TIGRESS]{Kim18}.}
\label{fig:sim_comparisons}
\end{figure}

While the challenges in dynamic range are  substantial, scientific access to ever-increasing computing power continues to push the boundaries of possibility. In this paper, we describe the Cholla Galactic OutfLow Simulations (CGOLS) project, a suite of simulations that attempts to tackle the problem of understanding galactic winds via a systematic approach. Beginning with a simple, analytically tractable feedback model for a galactic wind, we progressively increase the physical complexity of the feedback implementation and the included physics through each simulation in the suite, allowing us to investigate for each case the way in which the numerical results diverge from simpler theoretical expectations. In addition to a novel setup, a primary advancement the CGOLS suite contributes to the study of these winds is the extreme resolution made possible through the use of our GPU-based simulation code, Cholla \citep{Schneider15}. While we cannot yet achieve $1\,\mathrm{pc}$ resolution over a volume of $100\,\mathrm{kpc}^3$, the production simulations described in this paper each have a resolution of $\approx5\,\mathrm{pc}$ over a volume of $10\times10\times20\,\mathrm{kpc}^3$, which represents over an order-of-magnitude improvement as compared to any other isolated-galaxy simulations of these phenomena (see Figure~\ref{fig:sim_comparisons}). This vast resolving power allows us to tackle the problem of simulating galactic winds with increased fidelity and enables us to pay careful attention to the role that finite numerical resolution may play in our understanding of galactic winds via simulation.

Broadly, CGOLS consists of a set of global galaxy simulations designed to study the evolution of galactic outflows between radii of $1 --10\,\mathrm{kpc}$. The GPU-native nature of Cholla allows us to run on \textit{Titan}, as of 2017 November, the fastest computer available for public science in the United States and the fifth fastest in the world\footnote{www.top500.org}. The immense power of \textit{Titan} and the efficiency of Cholla make it straightforward to achieve resolutions for these isolated galaxy simulations that far outstrip those previously possible. In order to place these simulations in context, we plot in Figure~\ref{fig:sim_comparisons} a comparison between the total hydrodynamic volume elements (particles or cells) and resolution of the CGOLS suite and those of other simulations that have been used to study the structure and dynamics of galactic outflows. The relative locations of a representative sample of ISM patch and isolated galaxy simulations, zoom-in and static mesh refinement simulations, and cosmological simulations are outlined. As this plot demonstrates, the CGOLS project investigates winds in an entirely new region of parameter space, one in which the effects of global wind geometry and propagation into the CGM can be explored, while the resolution remains sufficient to better capture small-scale hydrodynamic and/or thermal instabilities that affect the multiphase structure of the outflow. \added{Essentially, we are applying the numerical resolution of the current premier cosmological simulations to the inner halo of a single galaxy.}

We organize this paper as follows. In Section~\ref{sec:simulations}, we provide details of the simulation suite, including the setup of the isolated galaxy and halo, the feedback model, and the relevant details of our code. In Section~\ref{sec:results}, we present results from the first simulation in the CGOLS suite, which investigates an adiabatic, nuclear starburst wind model. In Section~\ref{sec:xrays}, we connect this simulation to observations of the nuclear starburst galaxy, M82, with a focus on the observed diffuse soft X-ray emission. In Section~\ref{sec:convergence}, we explore the role of resolution and discuss convergence of the results. We conclude in Section~\ref{sec:conclusions}. A companion paper (Schneider et al., \textit{submitted}) presents results from simulations in the suite that include radiative cooling, and therefore investigates the potential multiphase nature of these outflows.

\section{Simulations}\label{sec:simulations}

Table~\ref{tab:simulations} lists the base set of simulations in the CGOLS suite. 
The scientific results presented in this paper primarily result from the first three simulations shown in the table, but Table~\ref{tab:simulations} includes additional CGOLS simulations for completeness. Each CGOLS simulation shares the same galaxy initial conditions and halo model, as described below.

\begin{deluxetable*}{lcccccc}
\tabletypesize{\scriptsize}
\tablecaption{Summary of current simulations in the CGOLS suite.\label{tab:simulations}}
\tablehead{
\colhead{Model} &
\colhead{$\mathrm{n}_x\times\mathrm{n}_y\times\mathrm{n}_z$\tablenotemark{a}} &
\colhead{$\mathrm{N}_\mathrm{cells}$\tablenotemark{b}} &
\colhead{$\mathrm{L}_x\times\mathrm{L}_y\times\mathrm{L}_z$ [kpc]\tablenotemark{c}} &
\colhead{$\Delta x$ [pc]\tablenotemark{d}} &
\colhead{Cooling} & 
\colhead{Feedback}
}
\startdata
A-2048 & $2048\times2048\times4096$ & $1.7\times10^{10}$ & $10\times10\times20$ & 4.88 & No & Central \\
A-1024 & $1024\times1024\times2048$ & $2.1\times10^{9}$ & $10\times10\times20$ & 9.77 & No & Central \\
A-512 & $512\times512\times1024$ & $2.7\times10^{8}$ & $10\times10\times20$ & 19.5 & No & Central \\
B-2048 & $2048\times2048\times4096$ & $1.7\times10^{10}$ & $10\times10\times20$ & 4.88 & Yes & Central \\
B-1024 & $1024\times1024\times2048$ & $2.1\times10^{9}$ & $10\times10\times20$ & 9.77 & Yes & Central \\
B-512 & $512\times512\times1024$ & $2.7\times10^{8}$ & $10\times10\times20$ & 19.5 & Yes & Central \\
C-2048 & $2048\times2048\times4096$ & $1.7\times10^{10}$ & $10\times10\times20$ & 4.88 & Yes & Clustered \\
C-1024 & $1024\times1024\times2048$ & $2.1\times10^{9}$ & $10\times10\times20$ & 9.77 & Yes & Clustered \\
C-512 & $512\times512\times1024$ & $2.7\times10^{8}$ & $10\times10\times20$ & 19.5 & Yes & Clustered \\
\enddata
\tablenotetext{a}{Number of cells in the $x$, $y$, and $z$ directions.}
\tablenotetext{b}{Total number of cells in the simulation.}
\tablenotetext{c}{Length of the simulation domain in physical units.}
\tablenotetext{d}{Physical resolution of the simulation (length of a cell).}
\tablecomments{This paper discusses results from the adiabatic central feedback simulations, A-2048, A-1024, and A-512. The remaining simulations in the suite are discussed in a companion paper (Schneider et al., \textit{submitted}).}
\end{deluxetable*}

\subsection{Simulation Setup and Galaxy Initial Conditions}

\begin{figure}
\includegraphics[width=\linewidth]{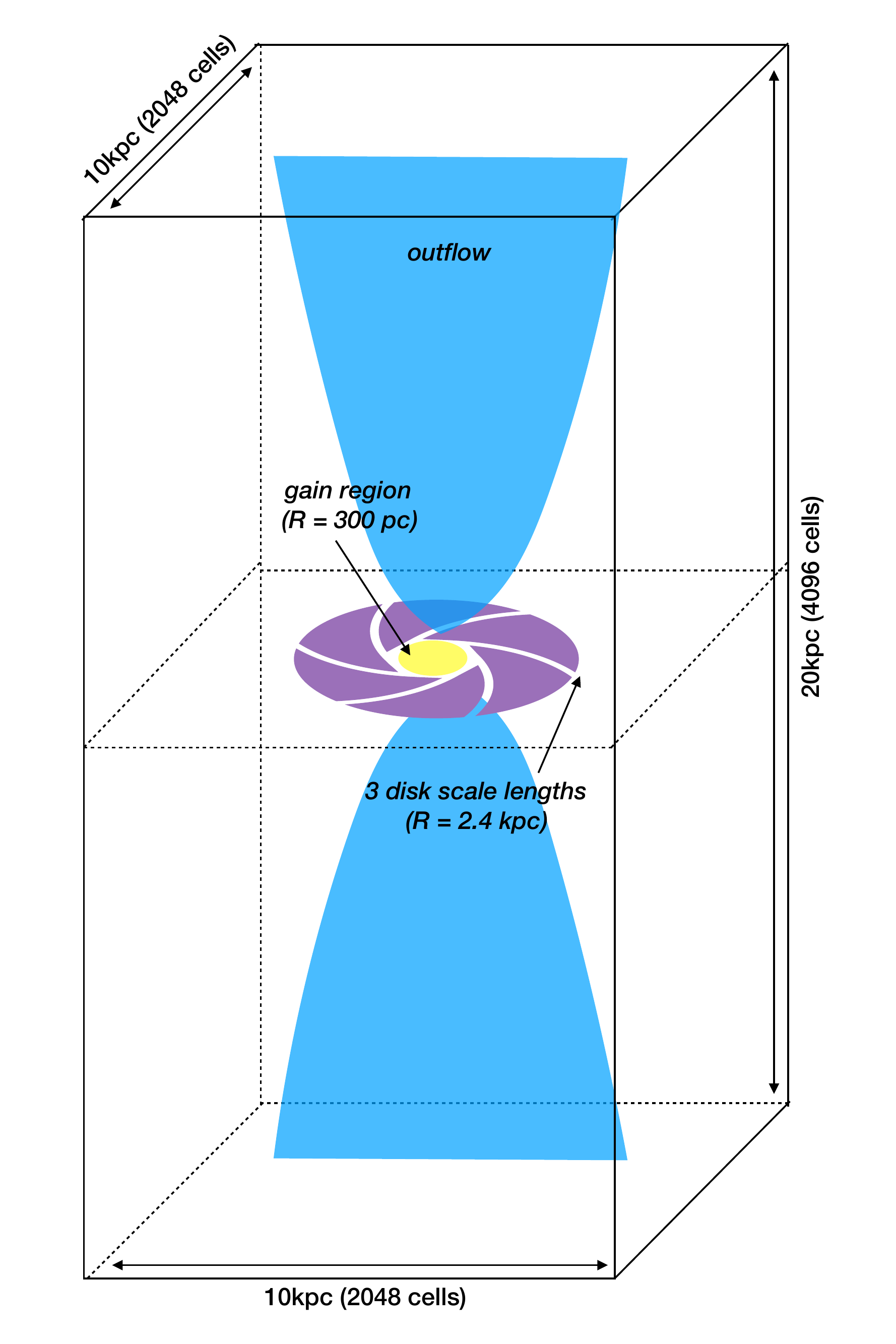}
\caption{Schematic illustrating the general setup of the CGOLS simulations.
Each simulation's initial conditions consist of a gaseous disk (purple disk), a dark matter halo, and a hot gaseous halo in initial hydrostatic equilibrium with the total gravitational potential. Within the disk is a ``gain region'' where energy and momenta from supernovae feedback are deposited (yellow region). The outflow resulting from the feedback then expands into the volume above and below the disk (blue regions).}
\label{fig:sim_schematic}
\end{figure}

A schematic representation of the simulation volume is shown in Figure~\ref{fig:sim_schematic}. Centered in the volume is a rotating, gas disk, initially surrounded by a static, adiabatic halo. The first feedback model we study in this project is simple and consists of a constant injection of mass and thermal energy into the central region of the galaxy, which drives a collimated wind. The fiducial simulation volume spans a region $10\times10\times20\,\mathrm{kpc}^3$, with $2048\times2048\times4096$ cells, giving a constant resolution of $\Delta x = 4.9\,\mathrm{pc}$ across the entire volume. Additional simulations are also performed at lower resolution (see Table~\ref{tab:simulations}).

\subsubsection{Gravitational Potential}
\begin{figure}
\includegraphics[width=\linewidth]{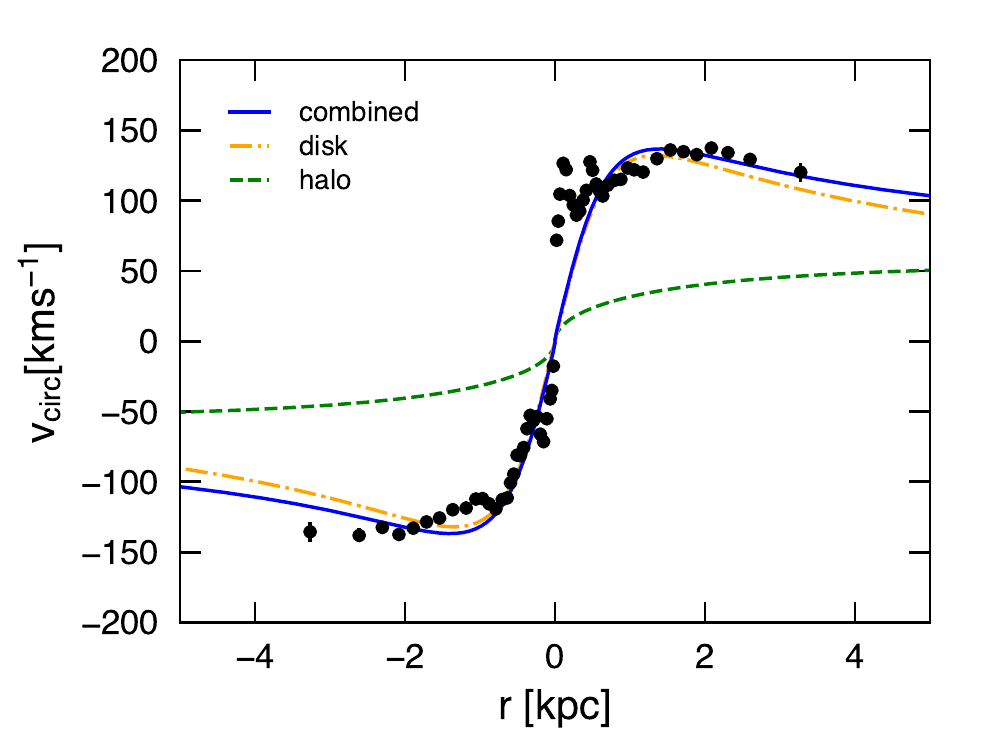}
\caption{Totation curve in the $x-y$ plane corresponding to the galactic potential of our simulated disk galaxy is shown with the solid blue line, while the dashed lines show separate disk and halo contributions. The black points are stellar rotation curve data from \cite{Greco12}. Our rotation curve is not a fit to these data, but rather represents the good agreement between our chosen simulation parameters and the observed dynamics of M82.}
\label{fig:rotation_curve}
\end{figure}

We use a static gravitational potential to model the contributions of both the stellar component of the galactic disk and dark matter halo throughout the simulations. Both the disk and halo are modeled after the gas-rich nearby starburst galaxy, M82. Self-gravity of the gas is not included in these initial CGOLS simulations. We use a Miyamoto-Nagai profile for the galaxy's stellar disk component \citep{Miyamoto75}, defined by
\begin{equation}
\Phi_\mathrm{stars}(r, z) = - \frac{G M_\mathrm{stars}}{\sqrt{r^2 + \left(R_\mathrm{stars} + \sqrt{z^2 + z_\mathrm{stars}^2}\right)^2}},
\end{equation}
where $r$ and $z$ are the radial and vertical cylindrical coordinates, $M_\mathrm{stars} = 10^{10}\,\mathrm{M}_\odot$ is the mass of the stellar disk \citep{Greco12}, $R_\mathrm{stars} = 0.8\,\mathrm{kpc}$ is the stellar scale radius \citep{Mayya09}, and $z_\mathrm{stars} = 0.15\,\mathrm{kpc}$ is the stellar scale height \citep{Lim13}. We use an NFW profile to represent the halo potential \citep{Navarro96}, defined by
\begin{equation}
\Phi_\mathrm{halo}(r) = -\frac{G M_\mathrm{halo}}{r[\mathrm{ln}(1 + c) - c/(1+c)]}\mathrm{ln}\left(1+\frac{r}{R_\mathrm{halo}}\right),  
\end{equation}
where $r$ is the radius in spherical coordinates, $M_\mathrm{vir} = 5\times10^{10}\,\mathrm{M}_\odot$ is the dark matter mass of the halo, $c = 10$ is the halo concentration, and $R_\mathrm{halo}$ is the scale radius of the halo, which is set by $R_\mathrm{h} = R_\mathrm{vir}/c = 5.3\,\mathrm{kpc}$
with our estimate of the virial radius, $R_\mathrm{vir} = 53\,\mathrm{kpc}$. We have estimated the dark matter mass and virial radius of the halo based on scaling relations between the stellar content of galaxies and their dark matter halos \citep[e.g][]{Kravtsov13}, but expect values within a factor of a few for the halo parameters to have little effect on the results of our simulations. The total gravitational potential in our simulations is then $\Phi = \Phi_\mathrm{stars} + \Phi_\mathrm{halo}$. Figure \ref{fig:rotation_curve} shows the theoretical stellar rotation curve in the $x-y$~plane of our simulated galaxy potential plotted against stellar data from M82 \citep{Greco12}. (The initial simulations in the CGOLS suite do not include star particles and only model gas. Correspondingly, Figure \ref{fig:rotation_curve} merely demonstrates that the adopted potential provides a good match to the M82 observations.)

\subsubsection{Disk Gas}
The galaxy itself in our simulations consists solely of gas.  The gas is distributed radially with an exponential surface density profile defined by the scale radius, $R_\mathrm{gas}~=~1.6\,\mathrm{kpc} = 2 \times R_\mathrm{stars}$, and total gas mass, $M_\mathrm{gas} = 2.5\times10^9\,\mathrm{M}_\odot$ \citep{Greco12}, according to
\begin{equation}
\Sigma(r) = \Sigma_0 \exp\left(-\frac{r}{R_\mathrm{g}}\right),
\end{equation}
\noindent
where $r$ is the cylindrical radial coordinate, and $\Sigma_0$ is the central surface density, $\Sigma_0 = M_\mathrm{gas}/(2 \pi R_\mathrm{gas}^2)$. The disk gas is isothermal with temperature $T_\mathrm{disk} = 10^4\,\mathrm{K}$, and is set in vertical hydrostatic equilibrium with the gravitational potential. This is achieved by solving for the vertical density distribution at each radial location,
\begin{equation}
\rho(z) = \rho_\mathrm{0,d} \exp\left[ -\frac{\Phi(z) - \Phi_\mathrm{0,d}}{c_\mathrm{s,d}^2}\right],
\end{equation}
where $\rho_\mathrm{0,d}$ is the midplane density, $\Phi_\mathrm{0,d}$ is the midplane potential, and $c_\mathrm{s,d}$ is the isothermal sound speed in the disk, $c_\mathrm{s,d}~=~\sqrt{k_\mathrm{B} T_\mathrm{disk} / \mu m_\mathrm{p}}$. When converting between mass density $\rho$ and number density $n$, we take $\mu = 0.6$ throughout the calculation, as appropriate for ionized gas with solar metallicity. The midplane density (at a given radius) is calculated by requiring the integrated density profile to equal the surface density, so
\begin{equation}
\rho_\mathrm{0,d} = \frac{\Sigma}{\int_{-\infty}^{\infty} \exp\left(-\frac{\Phi - \Phi_\mathrm{0,d}}{c_\mathrm{s,d}^2}\right)}.
\end{equation}

The disk gas is also in radial equilibrium with the static potential. Velocities are set by first calculating the tangential acceleration at a given radius due to the gravitational potential, then correcting this acceleration for the radial pressure gradient, i.e.
\begin{equation}
a_\phi(r, z) = -\nabla \Phi(r, z) + \frac{1}{\rho}\frac{\mathrm{d}P}{\mathrm{d}r},
\end{equation}
where $P = \rho c_\mathrm{s}^2 / \gamma$ is the gas pressure. We assume an adiabatic index $\gamma~=~5/3$ throughout the simulations. The disk gas is artificially truncated via an exponential ramp down in surface density beyond a radius of $R = 4.5\,\mathrm{kpc}$, in order to reduce possible boundary effects from the simulation box. 

\subsubsection{Halo Gas}
The isothermal gas disk is embedded in an adiabatic hydrostatic halo. To set the initial conditions for the halo gas, we first determine the density profile as a function of spherical radius, which for an adiabatic gas in hydrostatic equilibrium  with a spherical potential is given by
\begin{equation}
\rho(r) = \rho_\mathrm{0,h}\left[1+(\gamma-1)\frac{\Phi - \Phi_\mathrm{0,h}}{c_\mathrm{s, h}^2}\right]^\frac{1}{\gamma-1}.
\label{eqn:halo_profile}
\end{equation}
In this case, we normalize the profile by manually setting the density\added{, $\rho_\mathrm{0,h}$,} at \replaced{the cooling radius}{a radius of $100\,\mathrm{kpc}$}. We assume $\rho_\mathrm{0,h} = 3\times 10^3\,\mathrm{M}_\odot\,\replaced{\mathrm{kpc}^{3}}{\mathrm{kpc}^{-3}}$, or a number density of around $n \approx 10^{-3.5}\,\mathrm{cm}^{-3}$. \deleted{at a cooling radius of $r_\mathrm{cool} = 100\,\mathrm{kpc}$}. Similarly, $\Phi_\mathrm{0,h}$ is the gravitational potential at the cooling radius. Unlike the disk, in setting the halo gas distribution the potential is taken to be spherically symmetric (in other words, we evaluate the disk component of $\Phi$ at $r = 0$ for all $z$). Finally, the sound speed in Equation~\ref{eqn:halo_profile} is set by assuming a temperature of $T = 10^6\,\mathrm{K}$ at \replaced{the cooling radius}{$100\,\mathrm{kpc}$}. \added{The normalization of the number density and temperature we have chosen for the halo gas may be more appropriate for a slightly larger galaxy. However, these values for the M82 halo at $100\,\mathrm{kpc}$ are highly uncertain given its location within the larger M81 group. In practice, the halo gas gets blown out of the domain at early times in our simulations, so the exact normalization does not affect our results.} Having set the density, we then set the pressure of the gas adiabatically using $P = K \rho^\gamma$, where $K = c_\mathrm{s,h}^2 \rho_\mathrm{0,h}^{1-\gamma} / \gamma$. 

\begin{figure}
\includegraphics[width=\linewidth]{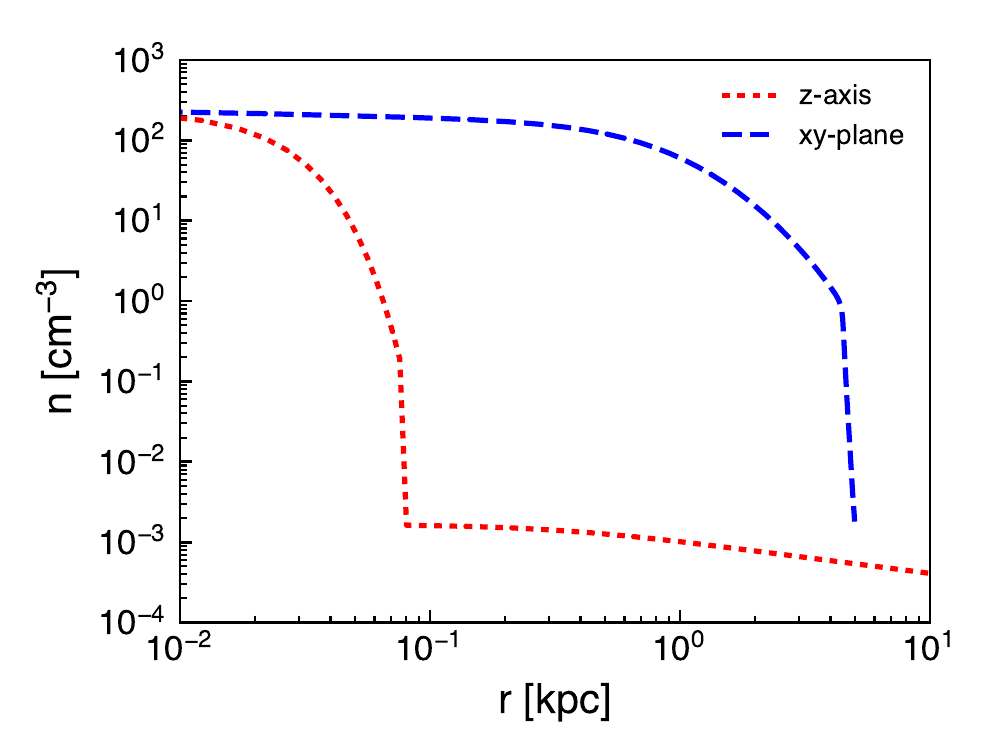}
\includegraphics[width=\linewidth]{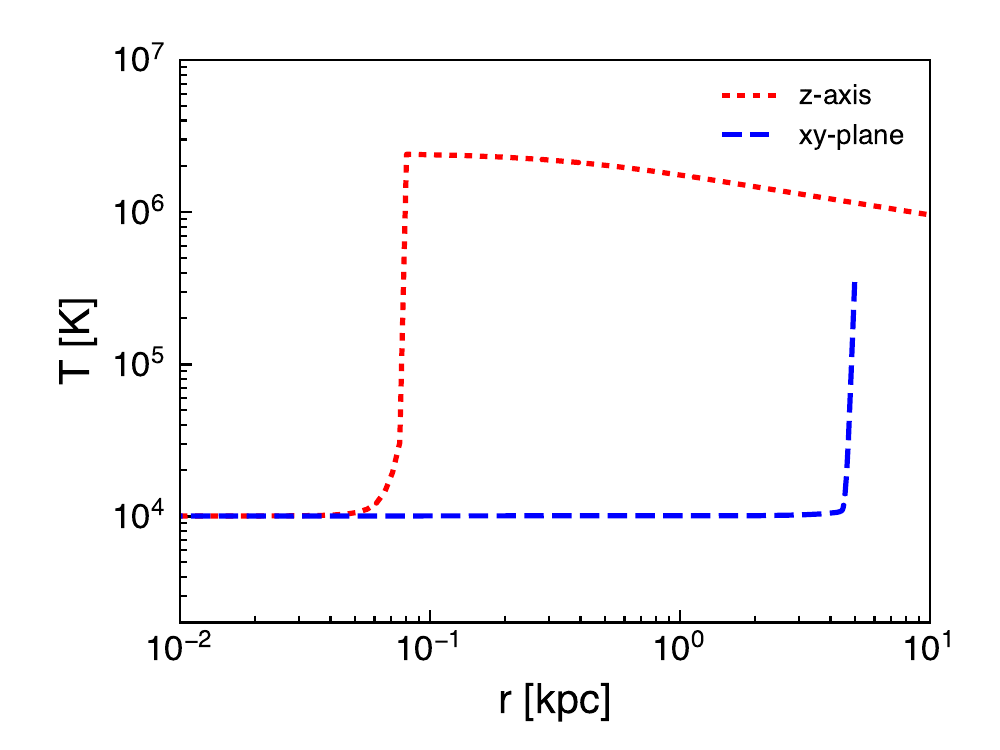}
\caption{Density and temperature profiles as a function of radius for our isothermal disk plus adiabatic halo galaxy initial conditions. Red dotted lines show the values along the $z$-axis, while blue dashed lines show the values in the disk midplane. Despite high surface densities in the center of the disk, our $<5\,\mathrm{pc}$ spatial resolution means that the disk scale height is still well resolved.}
\label{fig:ICs_profile}
\end{figure}

The density and internal energy from the halo profile are added to every cell in the simulation volume, including those in the disk, ensuring a smooth transition between disk and halo. Figure~\ref{fig:ICs_profile} shows the resulting one-dimensional density and temperature profiles as a function of radius along the $z$-axis and in the disk midplane. We have determined that these initial conditions are stable for over $1\,\mathrm{Gyr}$, well beyond the time scale of our galactic wind simulations. Shear forces at the interface between the rotating disk gas and the static gaseous halo do generate slight turbulence, but despite these perturbations the disk remains globally stable over long periods.

\pagebreak
\subsection{Feedback Model}

In this paper we will focus on the first, simplest feedback model that is implemented in the CGOLS suite. The series of simulations is designed to be systematic, building complexity in the feedback model from one simulation to the next, in order to better constrain our analytic models of feedback and winds. The first wind model we test is therefore that of a nuclear starburst driven by continuous star formation. The original analytic model for such a wind was developed by \cite{Chevalier85} (CC85) in order to explain the wind observed in the nearby starburst galaxy M82. \added{The CC85 model assumes adiabatic hydrodynamics and neglects the effects of gravity. We note that there have been several theoretical improvements to the CC85 model that include additional physics \citep{Wang95, Silich04, Thompson16, Bustard16}, which are addressed in our companion paper.}

In the CC85 model, a constant source of mass and energy are deposited in a spherical volume with radius $R$. The mass input rate $\dot{M}$ can be quantified as a function of the star formation rate, $\dot{M}_\mathrm{SFR}$, 
\begin{equation}
\dot{M} = \beta \dot{M}_\mathrm{SFR},
\end{equation}
where $\beta$ is known as the mass-loading factor of the wind. Similarly, the energy input rate $\dot{E}$ is quantified as a fraction of the energy input by supernovae \replaced{$\dot{E}_\mathrm{SFR}$}{$\dot{E}_\mathrm{SN}$},
\begin{equation}
\dot{E} = \alpha \dot{E}_\mathrm{SN},
\label{eqn:energy_loading_SN}
\end{equation}
where the energy-loading factor, $\alpha$ accounts for radiative losses in the ISM. With the assumption that each supernova releases $10^{51}\,\mathrm{erg}$ of energy, that there is one supernova per $100\,\mathrm{M}_\odot$ of star formation, \added{and that the star formation rate is in units of $M_\odot\,\mathrm{yr}^{-1}$,} the energy loading can be rewritten as a function of the star formation rate,
\begin{equation}
\dot{E} = 3\times10^{41}\,\mathrm{erg}\,\mathrm{s}^{-1}\, \alpha \dot{M}_\mathrm{SFR}.
\end{equation}

To replicate this model in the initial CGOLS simulations, we continuously inject mass and energy into a spherical volume in the center of the galaxy disk. We use a radius of $R = 300\,\mathrm{pc}$ for the gain region. Cells along the edge of the sphere are weighted so as to recover the correct total injection rate at low resolution and alleviate grid effects (we find the weighting unnecessary in our higher-resolution simulations). Over the course of the simulation, we vary $\alpha$, $\beta$, and the star formation rate (SFR) in order to probe different wind regimes. Although radiative losses will not be discussed in this work, different mass- and energy-loading factors lead to different theoretical predictions regarding the importance of radiative cooling in the wind. Varying $\alpha$, $\beta$, and SFR allows us to test different theoretical models within a single simulation. We refer to the feedback model described here as ``central'' in Table~\ref{tab:simulations}. Other feedback models are discussed in additional CGOLS papers, including a clustered model that distributes the mass and energy nonisotropically throughout the central disk region (Schneider et al. \textit{submitted.}). 

We start the simulations by running for $5\,\mathrm{Myr}$ with no feedback, primarily to allow cells at the disk-halo interface to equilibrate in radiative simulations. At $5\,\mathrm{Myr}$, we begin the mass and energy injection with $\dot{M} = 1.5\,\mathrm{M}_\odot\,\mathrm{yr}^{-1}$ and $\dot{E} = 1.5\times10^{42}\,\mathrm{erg}\,\mathrm{s}^{-1}$, corresponding to $\beta = 0.3$ and $\alpha = 1.0$ with an assumed star formation rate of $\dot{M}_\mathrm{SFR} = 5\,\mathrm{M}_\odot\,\mathrm{yr}^{-1}$. We refer to this throughout the paper as the low mass-loading state. Immediately after starting the feedback, we ramp up to a higher mass-loading state, with $\dot{M} = 12\,\mathrm{M}_\odot\,\mathrm{yr}^{-1}$ and $\dot{E} = 5.4\times10^{42}\,\mathrm{erg}\,\mathrm{s}^{-1}$, corresponding to $\beta = 0.6$ and $\alpha = 0.9$ with an assumed star formation rate of $\dot{M}_\mathrm{SFR} = 20\,\mathrm{M}_\odot\,\mathrm{yr}^{-1}$. 
The ramping is a linear function of time, e.g.
\begin{equation}
\dot{M} = M_1 + \frac{t}{t_\mathrm{ramp}}(M_2 - M_1)
\end{equation}
for the mass-loading, where $M_1$ is the low mass-loading state, $M_2$ is the high mass-loading state, $t_\mathrm{ramp} = 5\,\mathrm{Myr}$, and $t$ is the total time since the feedback began. We then keep the simulation in the high mass-loading state for $30\,\mathrm{Myr}$, before ramping back down to the low mass-loading state, e.g.
\begin{equation}
\dot{M} = M_1 + \frac{t - 40\,\mathrm{Myr}}{t_\mathrm{ramp}}(M_1 - M_2),
\end{equation}
again over the course of $5\,\mathrm{Myr}$. The remaining $30\,\mathrm{Myr}$ is run in the low mass-loading state, for a total simulation run time of $75\,\mathrm{Myr}$. 

\begin{figure}
\includegraphics[width=\linewidth]{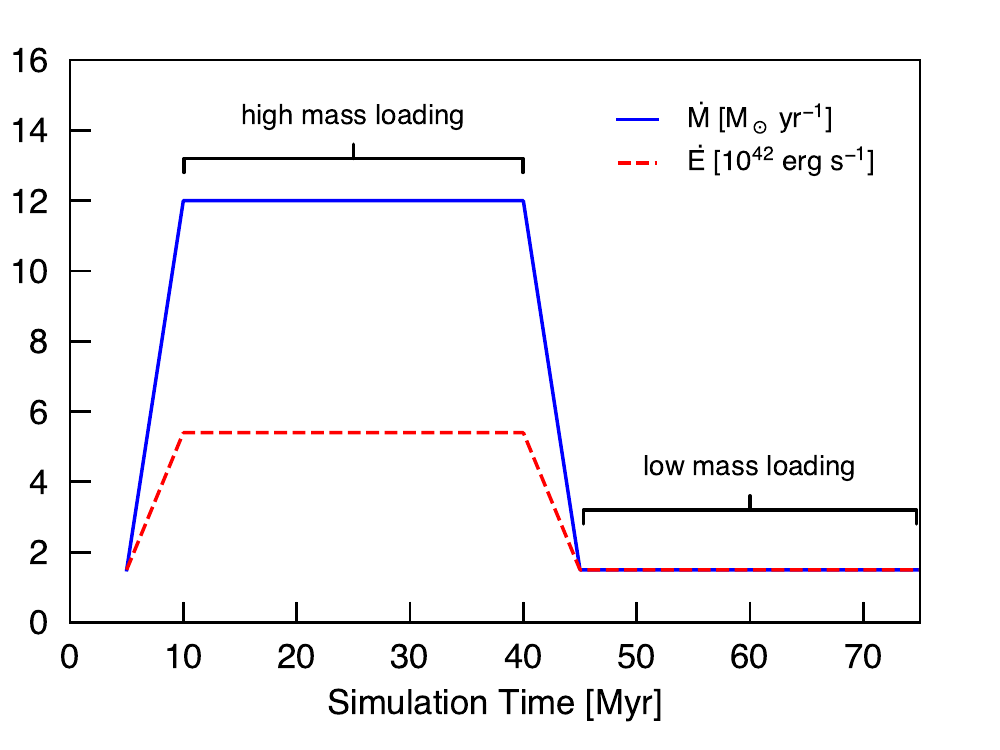}
\caption{Mass and energy loading in the simulation as a function of time. Blue and red lines plot the total injection rate of mass and energy, respectively. The outlined high mass-loading state corresponds to parameters $\beta = 0.6$, $\alpha = 0.9$, and $\mathrm{SFR} = 20\,\mathrm{M}_\odot\,\mathrm{yr}^{-1}$, while the low mass-loading state corresponds to $\beta = 0.3$, $\alpha = 1.0$, and $\mathrm{SFR} = 5\,\mathrm{M}_\odot\,\mathrm{yr}^{-1}$.}
\label{fig:feedback_params}
\end{figure}

\added{The mass and energy loading as a function of time are plotted in Figure~\ref{fig:feedback_params}. In addition to sampling an interesting region of parameter space, these $\alpha$, $\beta$, and SFR values are also physically motivated by the starburst itself. At the simulation start, we assume that some outside event (in this case, an interaction with nearby M81) has driven a large amount of gas to the center of the galaxy, triggering a burst of star formation. The mass loading may be higher at early stages in the burst because the supernova remnants are interacting with ISM gas that can add additional mass to the hot wind. Correspondingly, $\alpha$ is lower at early times as a result of radiative losses associated with these interactions \citep{Kim17}. As the burst matures, the star formation rate decreases, $\beta$ drops, and $\alpha$ increases as much of the ISM has been cleared out. The late-time low mass-loading parameters are directly motivated by constraints from X-ray observations of the hot wind in M82 today \citep{Strickland09}.}

\subsection{Hydrodynamical Model}

All of the simulations in the CGOLS suite are run using the Cholla hydrodynamics code \citep{Schneider15}, which includes a variety of spatial reconstruction techniques, Riemann solvers, and integrators. For this work, we employ piecewise linear reconstruction with limiting in the characteristic variables (PLMC), a linearized approximate Riemann solver that explicitly accounts for contact waves (HLLC), and a second-order predictor-corrector integration scheme. For simulations in the suite that require radiative cooling, we use an analytic approximation to the collisional ionization equilibrium (CIE) cooling curve that is applied via operator-splitting at the end of each hydrodynamic step. The static gravitational potential is also applied via operator-splitting. Each of these additions to the code is described in Appendix~\ref{app:cholla}. All boundaries of the simulation volume are set to allow gas to outflow only, i.e. we employ transmissive boundaries with a ``diode'' condition applied to the velocities \added{such that the conserved quantities (mass, momentum, and total energy) may exit but not enter the grid}. The Courant-Friedrichs-Lewy (CFL) number is set to 0.3 for all simulations.

\pagebreak
\section{Results from the Adiabatic Outflow Simulation}\label{sec:results}

In this section, we present scientific results from the first simulation carried out in the CGOLS suite, the adiabatic simulation A-2048. The most straightforward of the models, it consists of a centrally driven wind, and neglects the effects of radiative cooling. Thus, this model serves as a test of the original CC85 analytic outflow model within the context of a disk and halo and provides a basis of comparison for the more realistic models that are presented in other papers in this series.

In Figure~\ref{fig:slices}, we show density and temperature slices through the $x-z$ plane of the simulation at a series of characteristic times: 
$t=10$ Myr, when the initial superbubble is moving out into the CGM; 
$t=25$ Myr, about 15 Myr after the high mass-loading wind state has commenced; 
$t=50$ Myr, when the ramp down has occurred and the low mass-loading outflow solution is propagating outward; 
and $t=60$ Myr, about 15 Myr after the simulation reverts to the low mass-loading wind state. Relevant features of the outflow at each snapshot time include:

\begin{figure*}
\centering
\includegraphics[width=0.2254\linewidth]{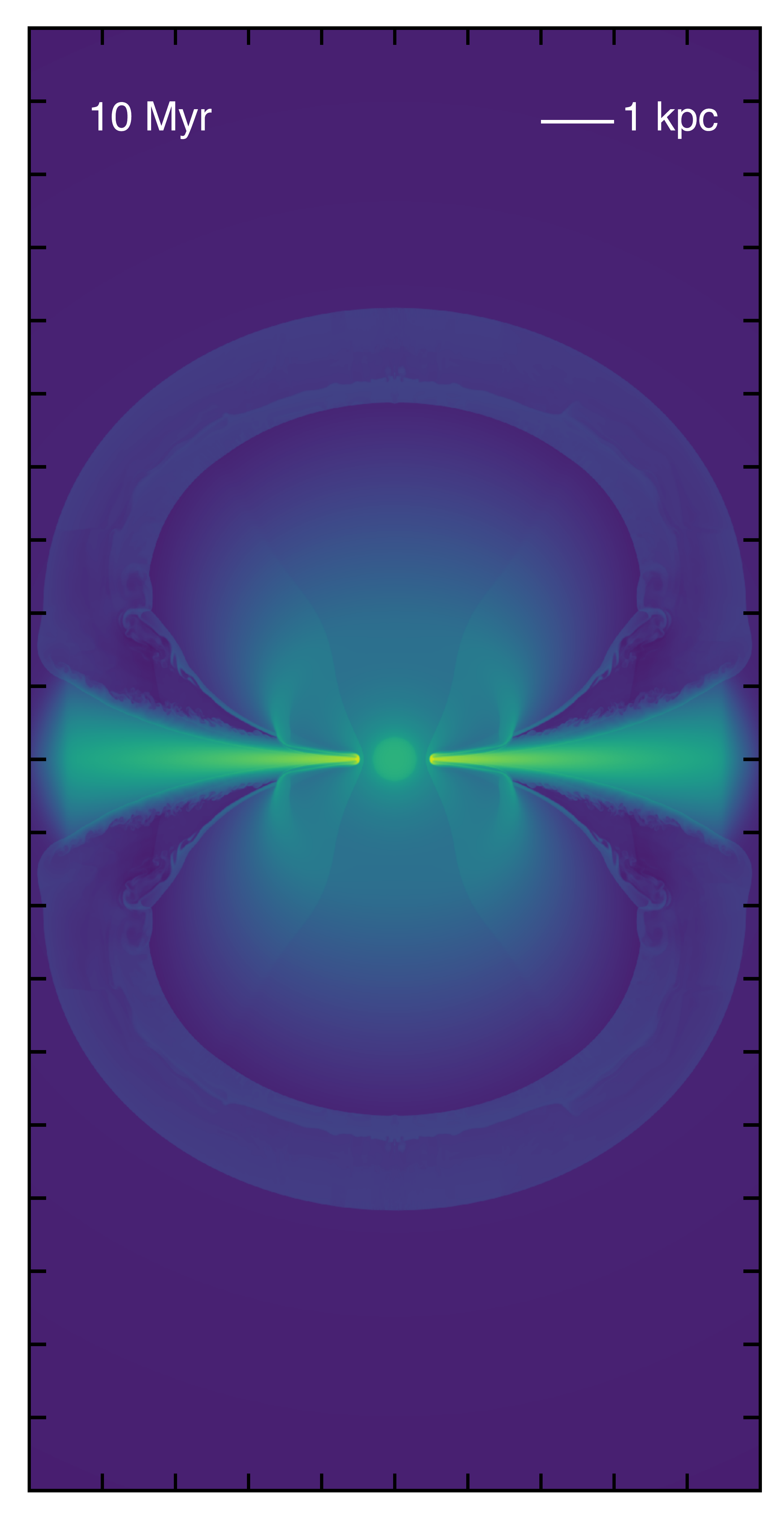}
\includegraphics[width=0.2254\linewidth]{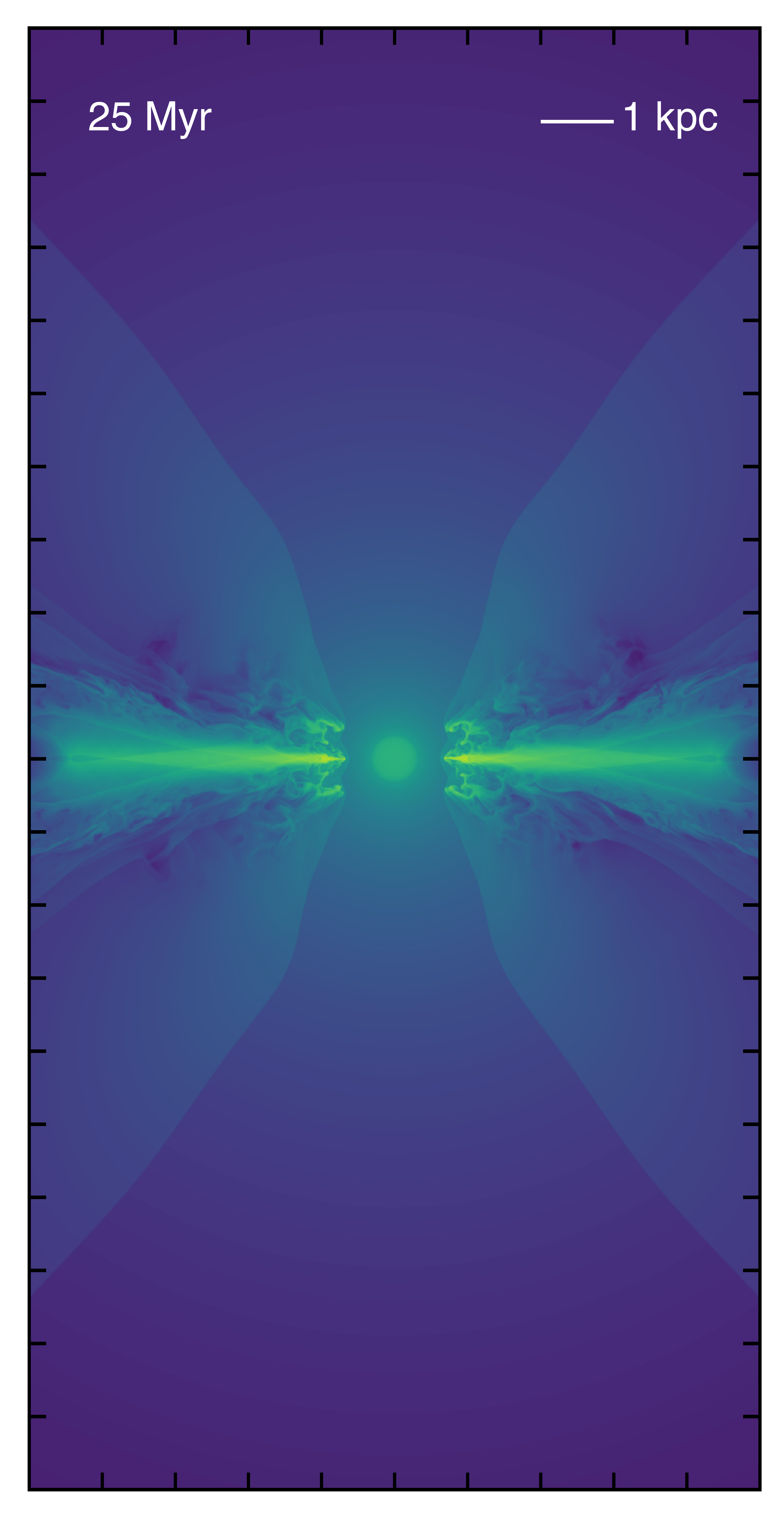}
\includegraphics[width=0.2254\linewidth]{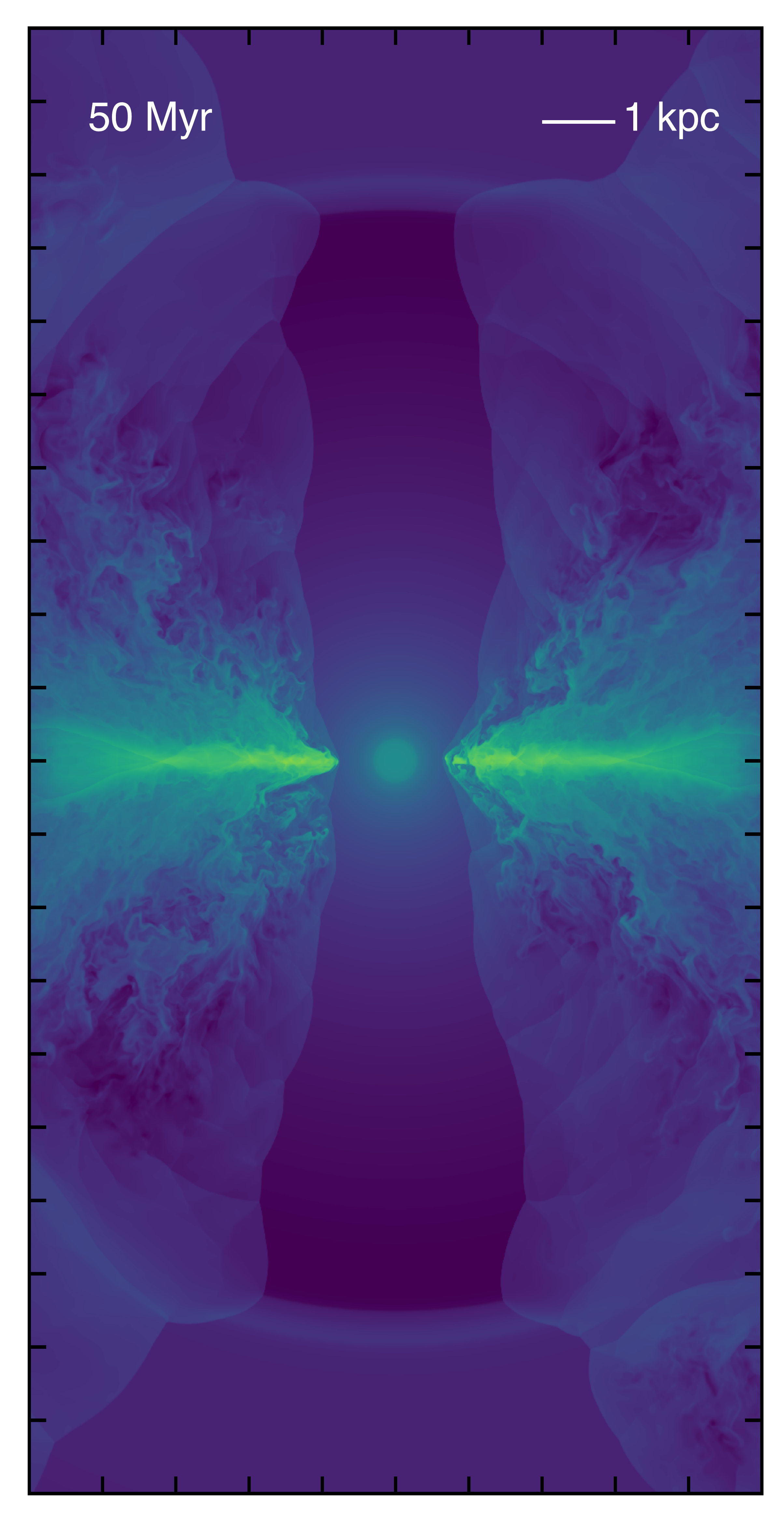}
\includegraphics[width=0.280\linewidth]{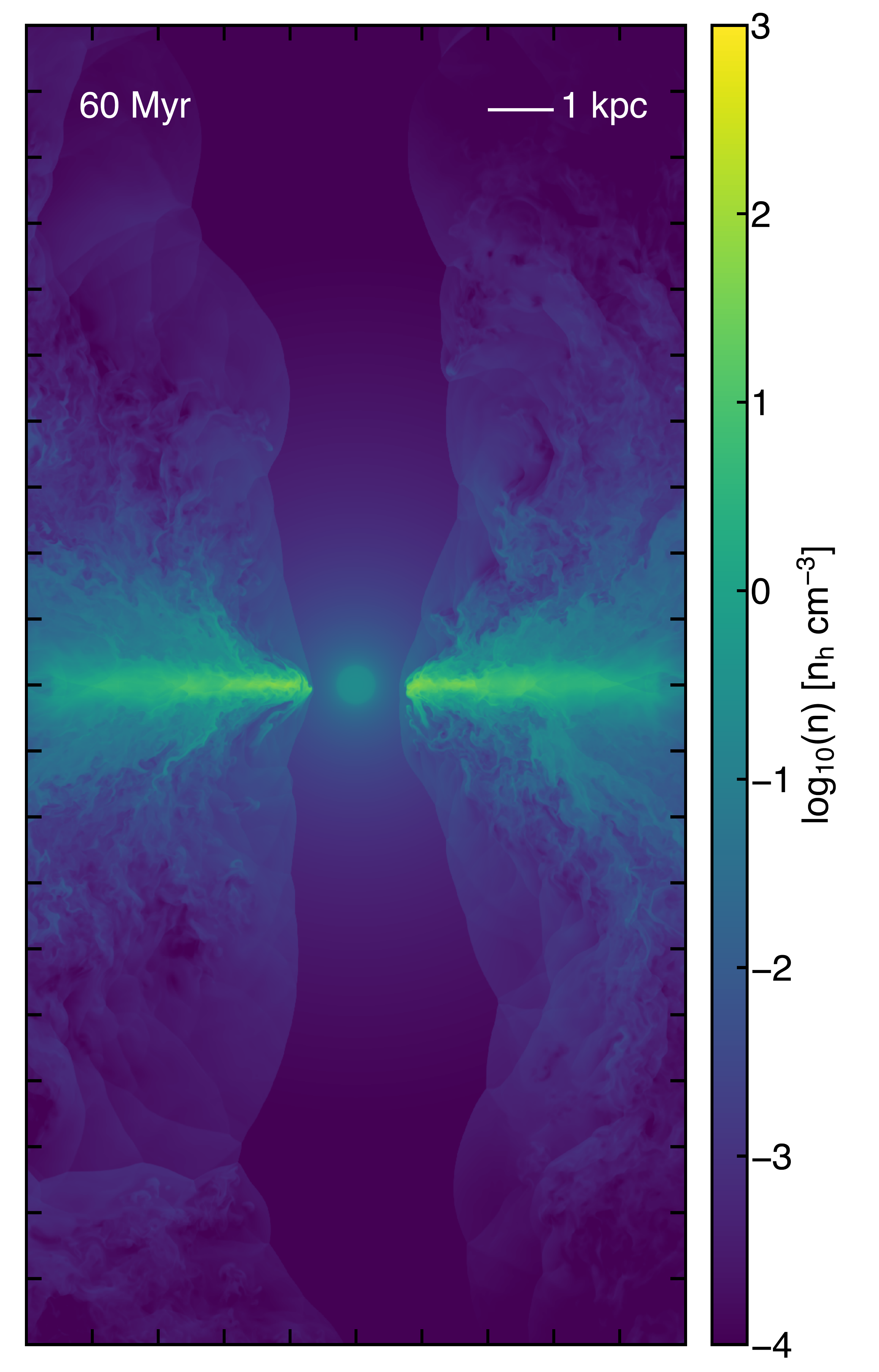}
\includegraphics[width=0.2254\linewidth]{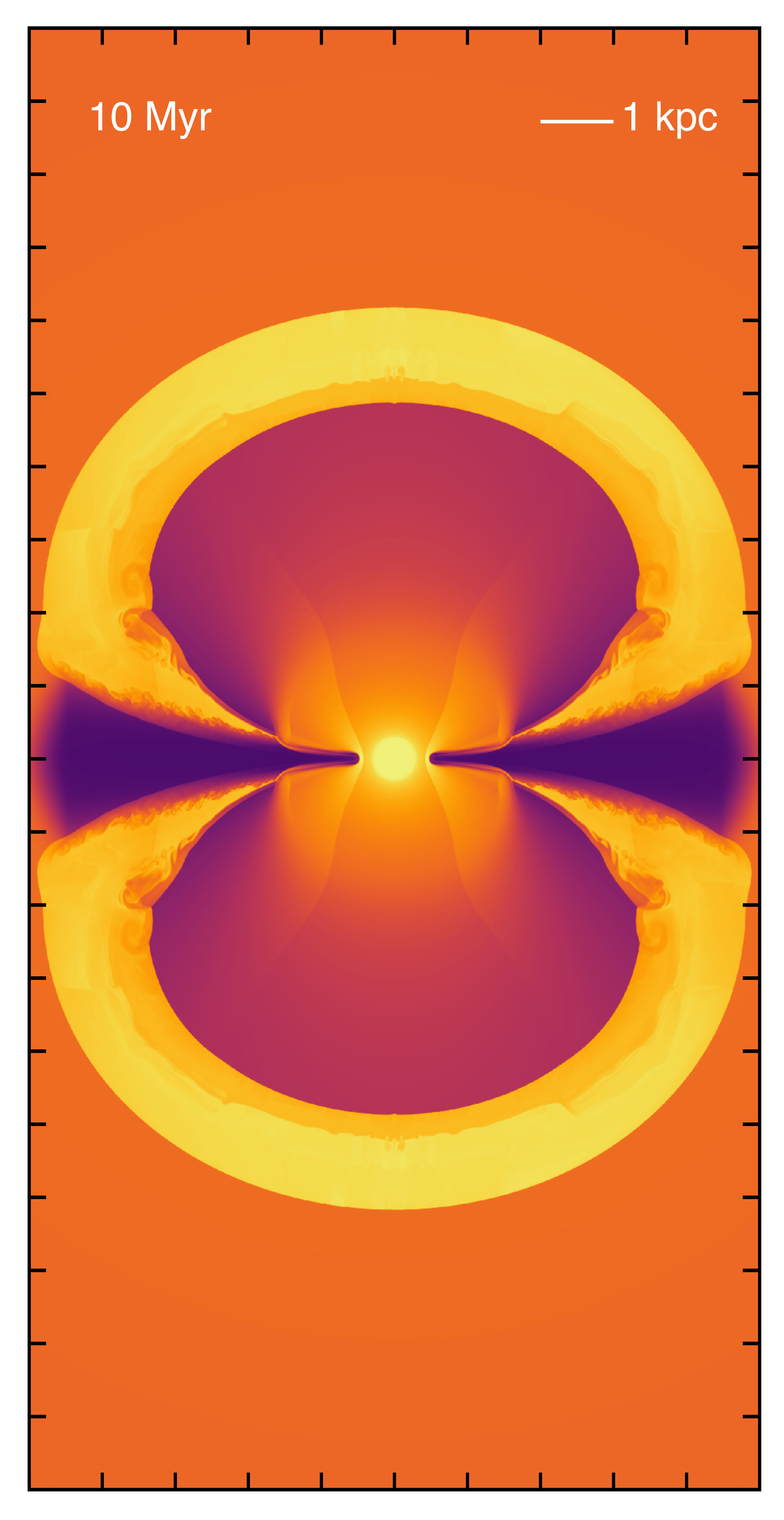}
\includegraphics[width=0.2254\linewidth]{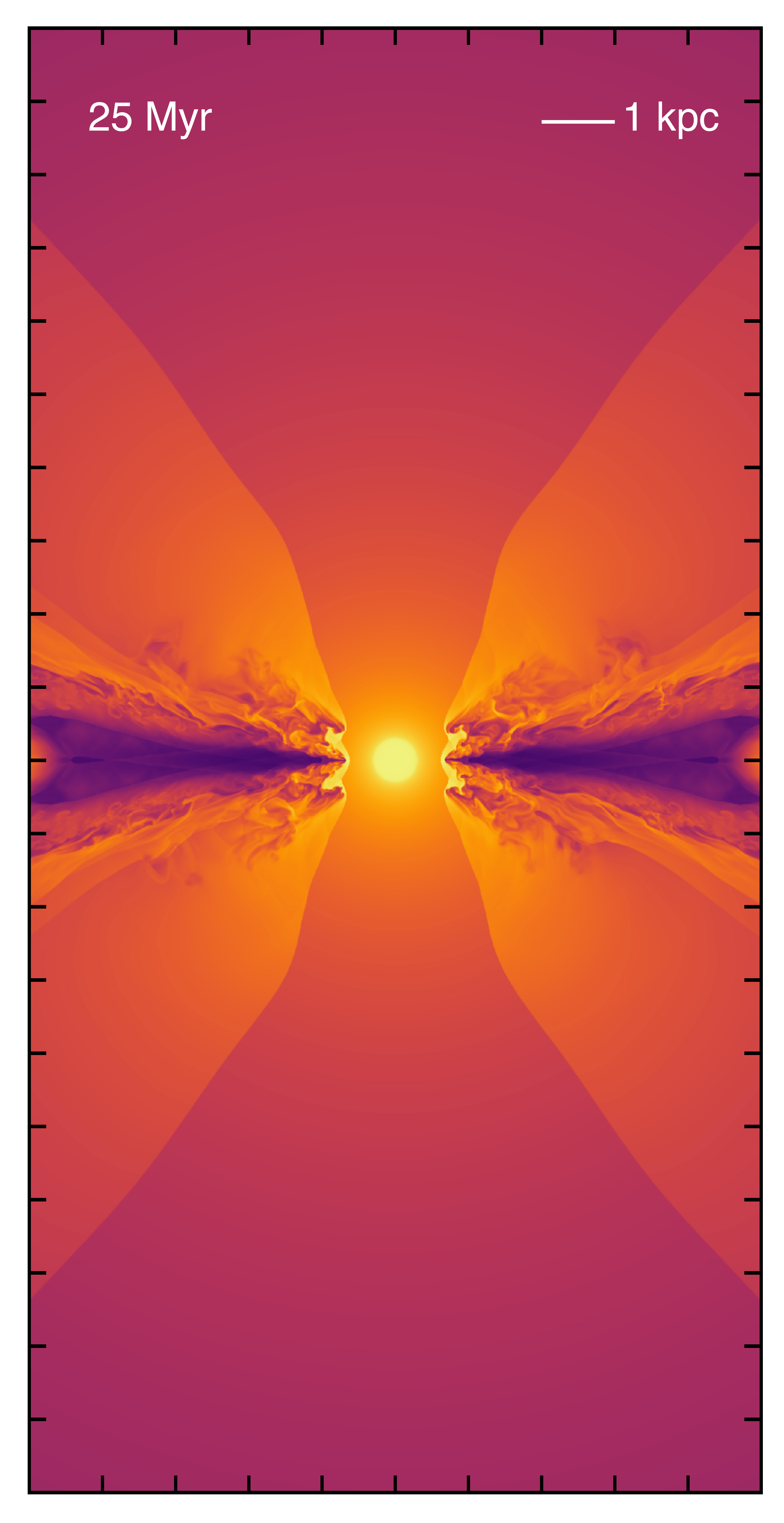}
\includegraphics[width=0.2254\linewidth]{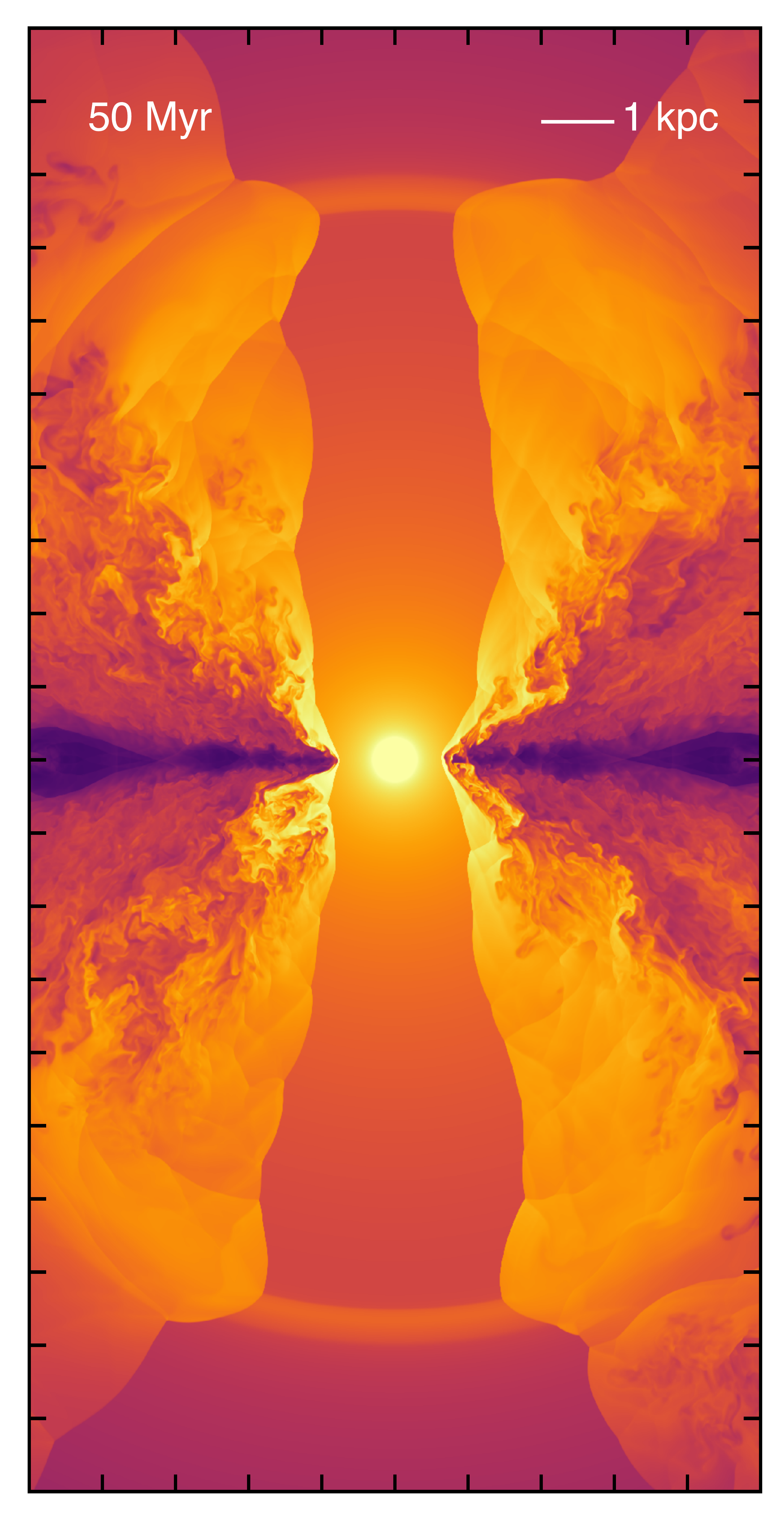}
\includegraphics[width=0.280\linewidth]{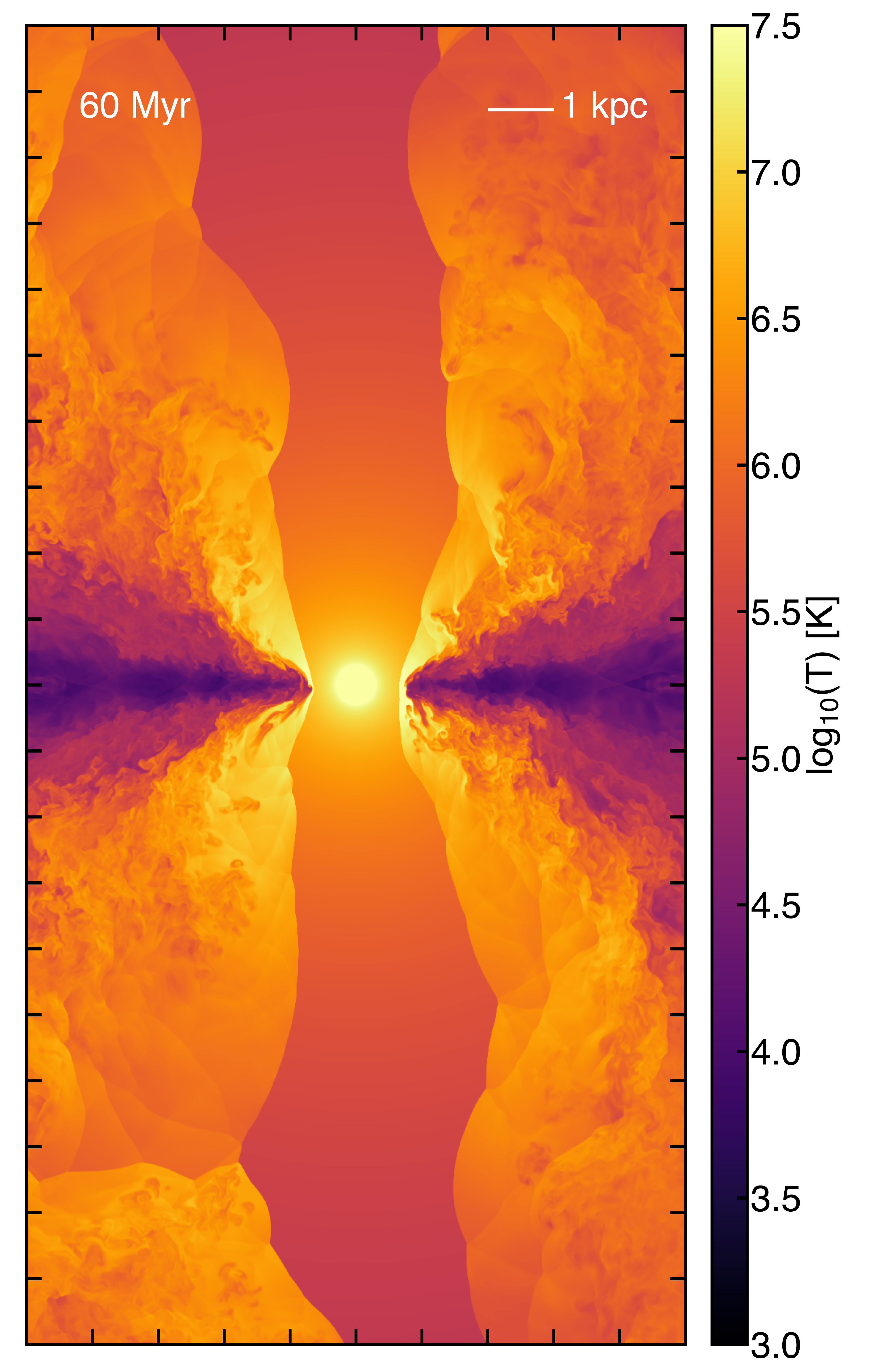}
\caption{Density and temperature slices through the $x-z$ plane of the adiabatic nuclear outflow simulation (model A-2048) at 10, 25, 50, and 60 Myr. The high-resolution ($\Delta x  = 4.9\,\mathrm{pc}$) and three-dimensional nature of the simulation results in a high degree of structure in the outflow, which is particularly visible in the turbulent regions at later times.}
\label{fig:slices}
\end{figure*}

\begin{itemize}
\item $t=10$ Myr: The superbubble created by the mass and energy input at the galaxy's center is propagating out into the CGM. The forward shock, contact 
discontinuity, and reverse shock are visible in both the density and temperature slice. Behind the reverse shock, the flow is smooth, and the beginnings of a pressure-confined conical outflow region can be seen. The isothermal disk remains relatively undisturbed at this time, though its central regions were blown 
out by the mass and energy injection immediately following the start of feedback. The outflow is just beginning to interact with the surface of the disk.

\item $t=25$ Myr: At this point, the high mass-loading outflow has stabilized in a biconical region centered on the $z$-axis. The free-wind cone is surrounded by a region of higher temperature and density caused by interactions between the outflow and disk. The wind from the central outflow has begun to eat away at the high-density gas near the center of the disk, driving turbulence along the interface, and the shock from the initial bubble has propagated through the disk, raising the temperature as compared to the initial conditions. This configuration remains more or less constant for the following $15\,\mathrm{Myr}$ of evolution in the high mass-loading state.

\item $t=50$ Myr: The simulation is now $5\,\mathrm{Myr}$ into the lower mass-loading state. While the outflow solution remained smooth during the ramp down from 40-45 Myr, the period $t=45-50$ Myr is characterized by a second outward-moving shock caused by the new outflow solution overtaking the old. The lower mass-loading state and higher $\alpha / \beta$ ratio results in a significantly higher terminal velocity, as expected from the analytic estimate,
\begin{equation}
v_\mathrm{term} = \sqrt{2\dot{E}/\dot{M}},
\end{equation}
which gives $v_\mathrm{term} = 1776\,\mathrm{km}\,\mathrm{s}^{-1}$ and  $v_\mathrm{term} = 1191\,\mathrm{km}\,\mathrm{s}^{-1}$ for the low and high mass-loading states, respectively. The turbulence at the interface has now increased considerably, and the resulting high pressure region combined with the lower pressure of the new outflow state is forcing the undisturbed outflow into an even narrower cone.

\item $t=60$ Myr: After $15\,\mathrm{Myr}$ in the low mass-loading state, the turbulent high pressure region from the interaction between the outflow and disk has overtaken most of the simulation volume. The undisturbed outflow now resembles a cylinder centered along the $z$-axis, and the disk itself has puffed up relative to the initial conditions as a result of the increased temperature and shocks that have now crossed it multiple times. Large clouds of higher-density disk gas continue to be ablated from the disk--wind interface, and are quickly shredded and carried away in the flow. The bow shocks from these interactions result in significantly higher temperature regions in the interface region than in the undisturbed outflow cone.
\end{itemize}

\subsection{Comparison with the Analytic Model}

Within the region of relatively undisturbed flow, we find an excellent fit between our simulations and the CC85 model. The CC85 model gives a solution for the spherically expanding flow in terms of the Mach number, which transitions from subsonic to supersonic at the edge of the mass and energy input region, $R$. The solutions for other physical variables of interest, such as density, velocity, and pressure, can be found as a function of radius by numerically integrating the spherical hydrodynamic equations with mass and energy source terms, e.g.:
\begin{equation}
\frac{1}{r^2}\frac{\mathrm{d}}{\mathrm{d}r}(\rho v_r r^2) = \frac{\dot{M}}{V},
\end{equation}
\begin{equation}
\rho v_r \frac{\mathrm{d}v_r}{\mathrm{d}r} = -\frac{\mathrm{d}P}{\mathrm{d}r}-\frac{\dot{M}}{V}v_r
\end{equation}
\begin{equation}
\frac{1}{r^2}\frac{\mathrm{d}}{\mathrm{d}r}\left[\rho v_r r^2\left(\frac{1}{2}v_r^2 + \frac{\gamma}{\gamma-1}\frac{P}{\rho}\right)\right] = \frac{\dot{E}}{V},
\end{equation}
where $v_r$ is the radial velocity, $V = \frac{4}{3}\pi R^3$ is the volume within the injection region, and all other symbols are as previously defined.

\begin{figure}
\includegraphics[width=1.0\linewidth]{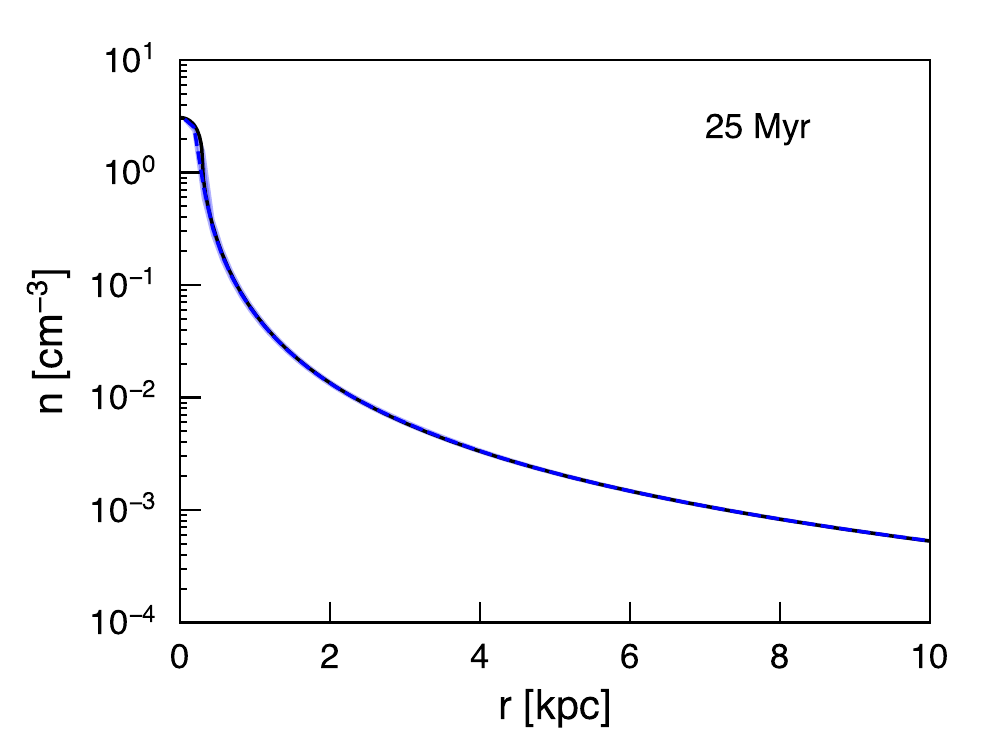}
\includegraphics[width=1.0\linewidth]{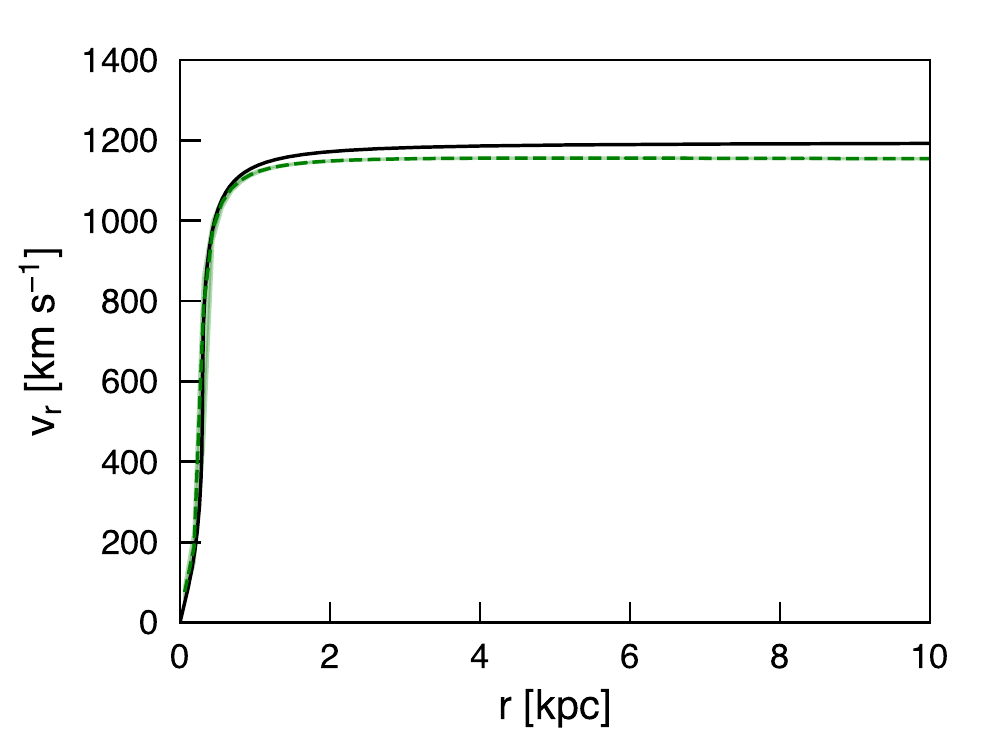}
\includegraphics[width=1.0\linewidth]{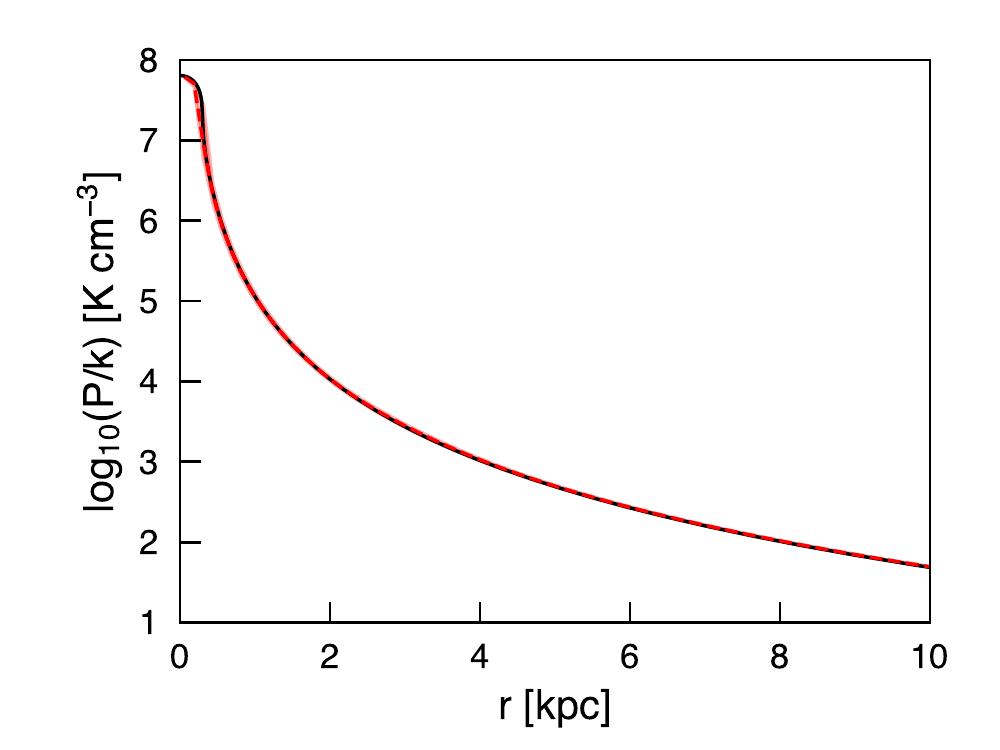}
\caption{Density, radial velocity, and pressure as a function of spherical radius are plotted within a biconical region with opening angle $\Delta\Omega = 60^\circ$. The analytic solution is shown with a black solid line in each panel, while the volumetric median of the values of the simulation data is plotted with a dashed line. \added{The mean, 25th, and 75th percentiles of the data within the cone are identical to the median (indicating that all the gas within the cone has the same fluid values)}. Data are from the $t=25\,\mathrm{Myr}$ snapshot, during the high mass-loading outflow state.}
\label{fig:chevalier_early}
\end{figure}

\begin{figure}
\includegraphics[width=1.0\linewidth]{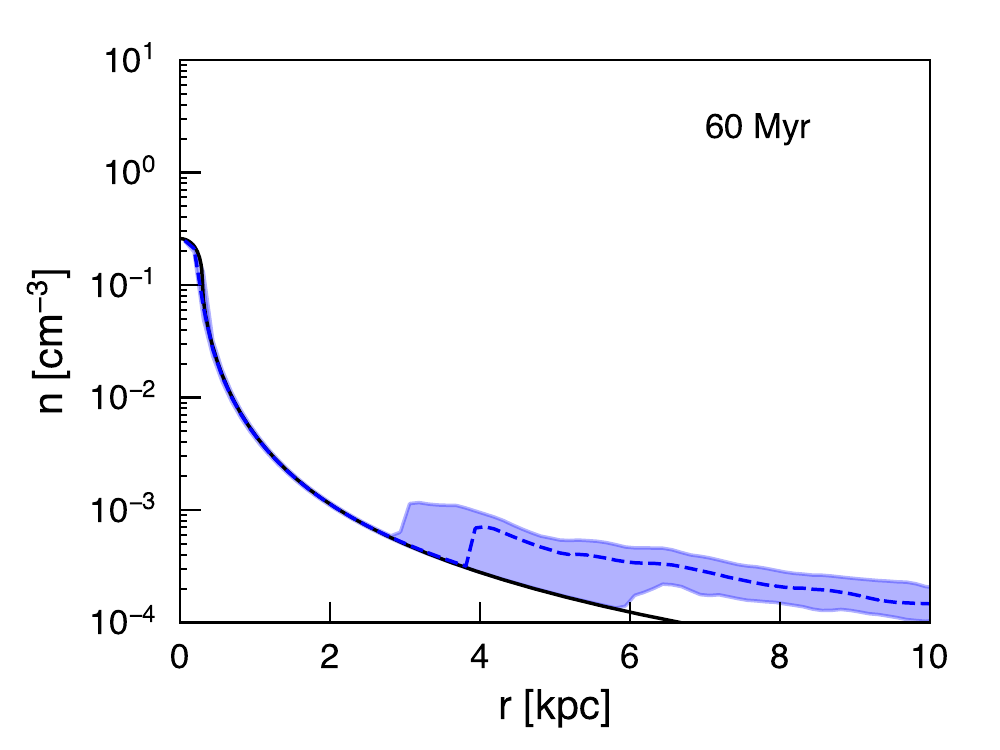}
\includegraphics[width=1.0\linewidth]{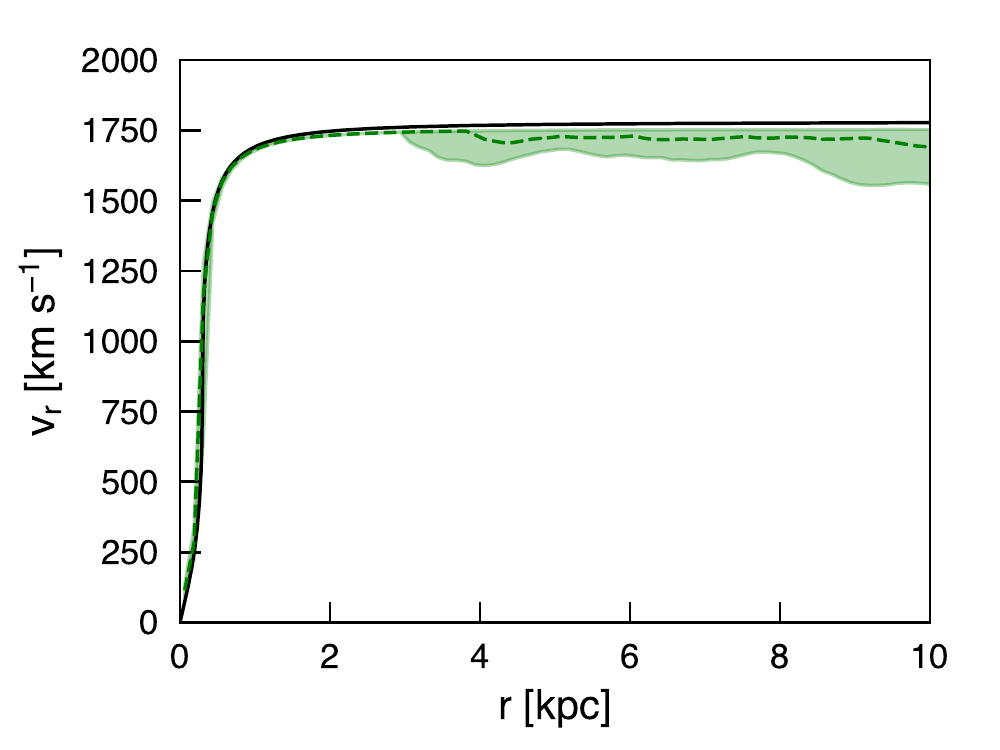}
\includegraphics[width=1.0\linewidth]{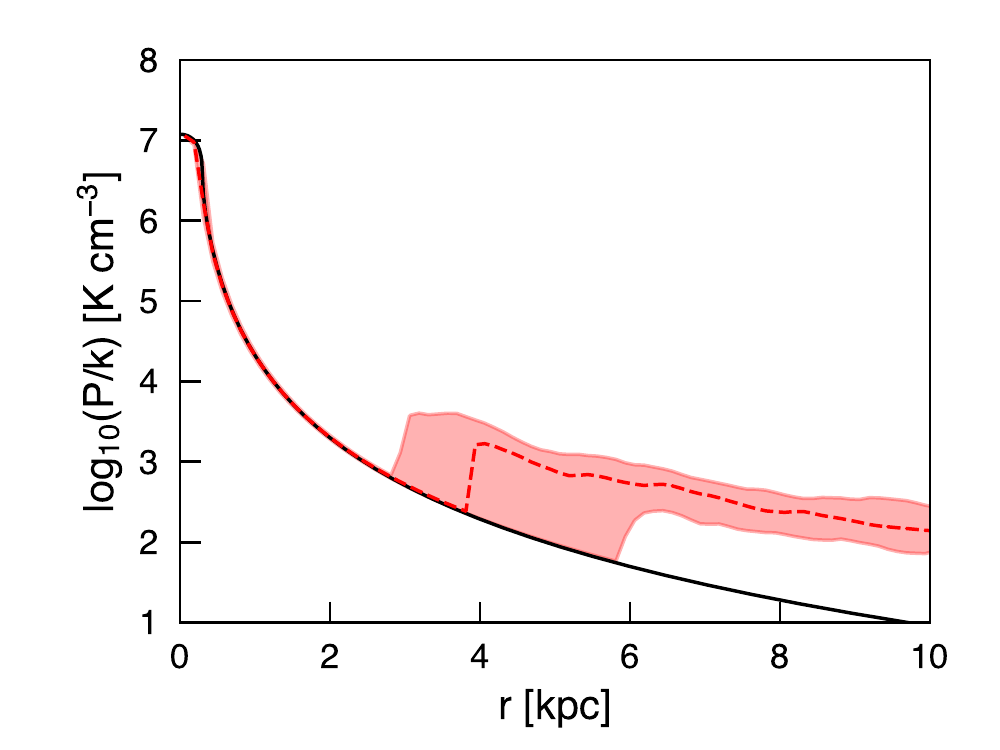}
\caption{Same as Figure~\ref{fig:chevalier_early}, but at $t=60\,\mathrm{Myr}$, during the low mass-loading outflow state. \added{The median values are shown with the dashed line, while the filled area represents all data in the outflow bicone between the 25th and 75th percentiles to give a sense of the variation within the region}. The sudden change in the character of the data at $r\sim3\,\mathrm{kpc}$ is a result of the conical selection region including some of the turbulent interface region at larger radii.}
\label{fig:chevalier_late}
\end{figure}

In Figure~\ref{fig:chevalier_early}, we compare the analytic solution predicted by the CC85 model to the results from the simulation at $t=25\,\mathrm{Myr}$ (in the high mass-loading state) and in Figure~\ref{fig:chevalier_late} at $60\,\mathrm{Myr}$ (in the low mass-loading state). Panels show the density, radial velocity, and pressure, with the analytic solution plotted as a black line. In order to focus the results on the outflow, only simulation data within a biconical region with an opening angle of $\Delta\Omega = 60^\circ$ centered on the $z$-axis are included in this analysis. Simulation data are binned as a function of spherical radius into 93 evenly spaced bins with $\Delta r = 0.125\,\mathrm{kpc}$.

At early times, the simulation data within the outflow cone match the analytic solution nearly exactly. The maximum velocity is slower by about $40\,\mathrm{km}\,\mathrm{s}^{-1}$, which can be attributed to the effect of the gravitational potential that is not taken into account in the analytic model. At later times, the results from the simulation appear more complex. The outflow cone during the low mass-loading state is narrower and less time-stable, effectively ``wobbling'' around as the pressure balance with the turbulent interface region changes. As a result, much of the volume within the measured biconical region is now filled with turbulent interface material. The turbulent region is characterized by gas that has higher density and lower velocity, leading to an overall increase in pressure and temperature, as seen at larger radii in Figure~\ref{fig:chevalier_late}. The shocks created by slowing the hot wind gas as it interacts with clouds of denser, cooler gas can explain this increased temperature. 
For example, at $r=10\,\mathrm{kpc}$, the mean radial velocity of the interface material measured within 
the $\Delta\Omega = 60^\circ$ bicone is $v_r = 1610\,\mathrm{km}\,\mathrm{s}^{-1}$, \replaced{a factor of}{approximately} $150\,\mathrm{km}\,\mathrm{s}^{-1}$ less than the terminal velocity predicted by the analytic model. Slowing the gas by this amount would increase its temperature by $\Delta T \sim 9 \times 10^5\,\mathrm{K}$, and indeed, we measure a mean temperature for the gas at $r=10$ kpc of $T = 1.05\times10^6\,\mathrm{K}$, while the analytic expectation for the outflow temperature at $r=10\,\mathrm{kpc}$ is $T\sim 10^5\,\mathrm{K}$. This slowing of the outflowing material and increased density and temperature should also affect the observable signature of soft X-ray emission, which we investigate in the following section.

\section{Soft X-Ray Emission}\label{sec:xrays}

Many studies of the soft X-ray emission have been conducted for M82 \citep[e.g.][]{Griffiths00, Strickland04, Li13}, one of the best-resolved starburst galaxies in the local universe and a unique target for investigating the spatial distribution of diffuse soft X-ray emission. We compare the soft X-ray emission from our adiabatic simulation to observations of the emission in M82, in order to both validate our model and provide additional insight as to the source of the observed emission. We focus our comparison on a snapshot from the simulation at $t=60\,\mathrm{Myr}$, as the later stages of the simulation represent the most relevant comparison to the present-day M82 system. The mass and energy loading during the late-time evolution of the simulation are set to match those of M82 by construction, based on estimates of $\alpha$ and $\beta$ made by studying the diffuse hard X-ray emission in the nuclear region \citep{Strickland09}. However, those constraints do not necessarily ensure a good match to the soft X-rays.

\begin{figure}[h!]
\includegraphics[width=1.0\linewidth]{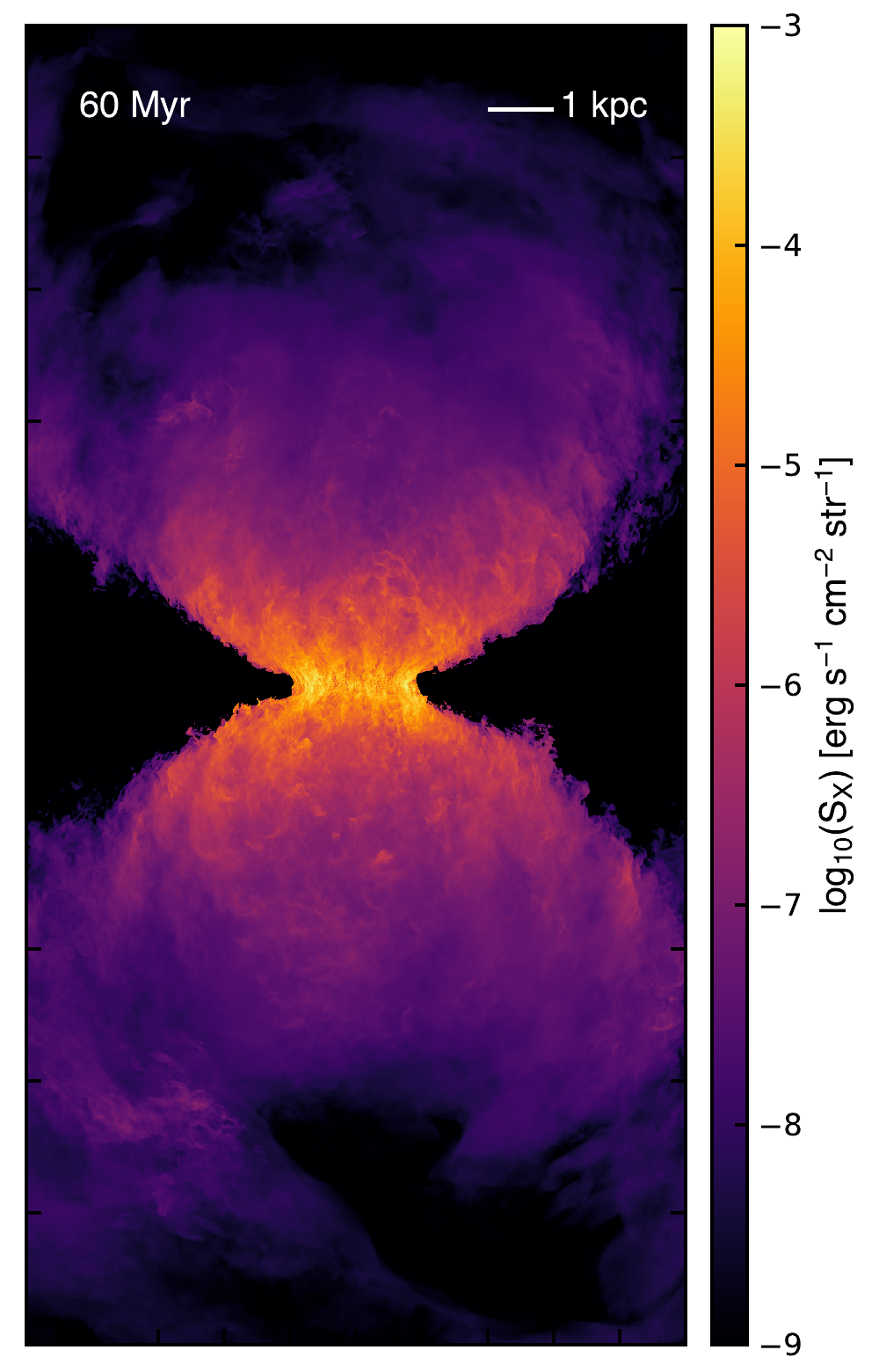}
\caption{Soft X-ray surface brightness at $60\,\mathrm{Myr}$ projected in the $x-z$ plane. The brightness is estimated by integrating the cooling rate along the line of sight of any cells with a temperature within the 0.3 - 2.0 keV range (see Equation~\ref{eqn:xray_brightness}).}
\label{fig:xrays}
\end{figure}

In Figure~\ref{fig:xrays} we show a map of the soft X-ray surface brightness at $t=60$ Myr. The map was made by integrating the estimated soft X-ray emission along the $y$-axis, where only cells with temperatures between 0.3 and $2.0\,\mathrm{keV}$ were included in the integral, i.e.
\begin{equation}
S_\mathrm{X} = \frac{1}{4\pi}\int_{0}^{L_y} n^2\Lambda(T) \mathrm{d}y \; [\mathrm{erg}\,\mathrm{s}^{-1}\,\mathrm{cm}^{-2}\,\mathrm{str}^{-1}].
\label{eqn:xray_brightness}
\end{equation}
We convert from temperature in a given cell to energy by assuming $E = \frac{3}{2}k_\mathrm{B} T$, as appropriate for a fully ionized plasma. The cooling curve used to estimate the emission, $\Lambda(T)$, is an analytic fit to a solar-metallicity collisional ionization equilibrium curve calculated with Cloudy \citep{Ferland13}, and is described in Appendix~\ref{app:cooling}. The line-of-sight integral gives a brightness in $\mathrm{erg}\,\mathrm{s}^{-1}\,\mathrm{cm}^{-2}$, which we convert to surface brightness by dividing by $4\pi$ steradian, yielding the values shown in the map in Figure~\ref{fig:xrays}. The $0.3 - 2.0\,\mathrm{keV}$ band was chosen for comparison to the \textit{Chandra} X-ray telescope ACIS data.

Overall, the simulated X-ray data compare favorably to the observations in both spatial extent and luminosity \citep[see e.g. the surface brightness map in Figure 2h of][]{Strickland04}. 
The integrated soft X-ray luminosity across the entire simulation volume is $L_\mathrm{X} = 1.9\times 10^{40}\,\mathrm{erg}\,\mathrm{s}^{-1}$. \cite{Strickland04} calculate a total soft X-ray luminosity of $L_\mathrm{X} = 4.3\times 10^{40}\,\mathrm{erg}\,\mathrm{s}^{-1}$, of which $L_\mathrm{X} = 2.3\times 10^{40}\,\mathrm{erg}\,\mathrm{s}^{-1}$ comes from the nuclear region, a circular aperture $1\,\mathrm{kpc}$ in radius. Given that this simulation is adiabatic, the extent to which the soft X-ray emission matches the observations should perhaps be surprising, but we do note that in this region of $\alpha$ and $\beta$ parameter space, we do not expect significant radiative losses from the outflow.

\begin{figure}
\includegraphics[width=1.0\linewidth]{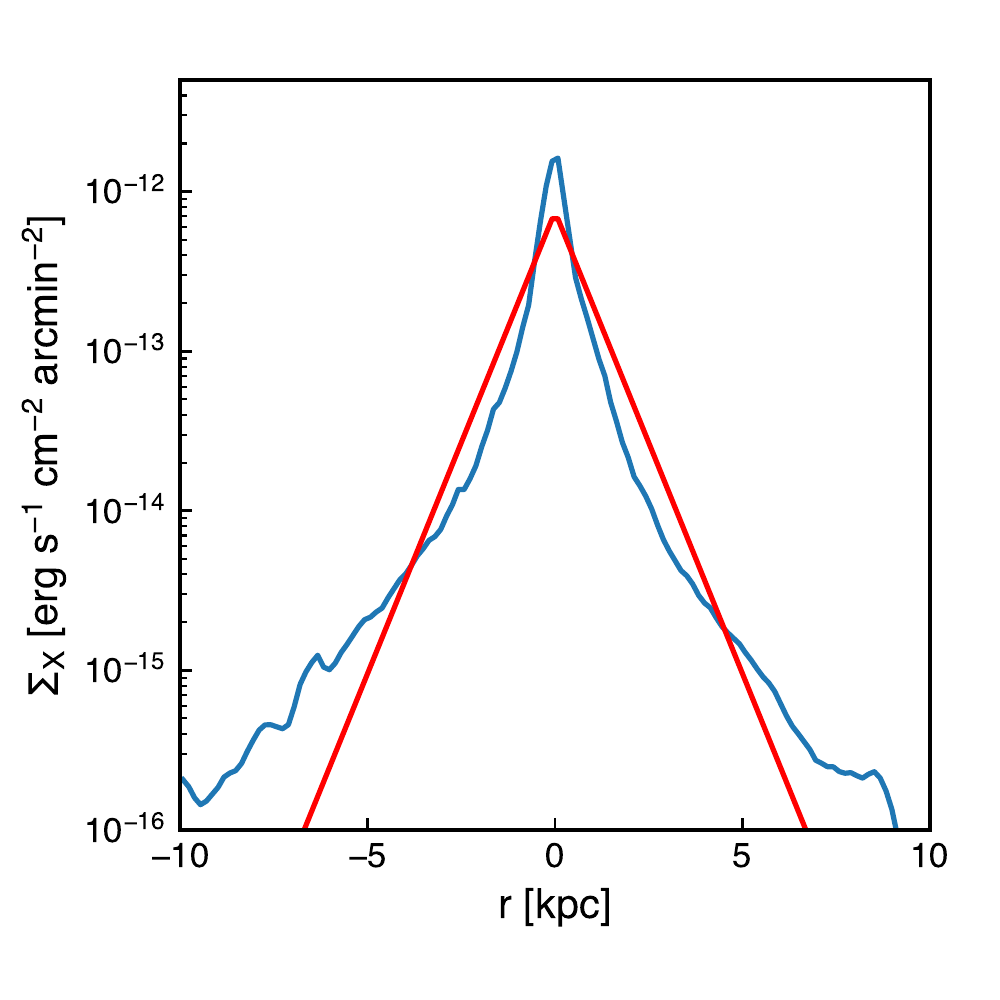}
\caption{Soft X-ray surface brightness profile as a function of $z$-height at 60 Myr. The X-ray luminosity is calculated as in Figure~\ref{fig:xrays}, and now integrated across the $x$-axis ($\Delta x = 10\,\mathrm{kpc}$), in slices with $\Delta z = 0.15625$. The exponential fit to M82 from \cite{Strickland04} is plotted in red.}
\label{fig:xray_profile}
\end{figure}

For a more quantitative look at how the spatial extent of the soft X-rays compares with the data, we plot in Figure~\ref{fig:xray_profile} the vertical surface brightness profile of the map shown in Figure~\ref{fig:xrays}. We integrate the surface brightness as calculated in Equation~\ref{eqn:xray_brightness} in slices with $\Delta x = 10\,\mathrm{kpc}$ and $\Delta z = 0.15625$, and then divide over the area to keep the units of $\mathrm{erg}\,\mathrm{s}^{-1}\,\mathrm{cm}^{-2}\,\mathrm{arcmin}^{-2}$ (with a conversion from steradians to arcminutes). We have also plotted the best-fit exponential profile from \cite{Strickland04}, which was calculated in the same manner. To facilitate comparison with the observations, we have converted \textit{Chandra} counts in the $0.3 - 2.0\,\mathrm{keV}$ band to energies assuming $1\,\mathrm{photon}\sim 2\times 10^{-9}\,\mathrm{erg}$ \citep[see e.g. Section 4.3.2 of][]{Strickland04}.

Again, we find a good match between the simulated data and the observations. The extended filamentary structures seen in the simulation lead to a surface brightness profile that is less peaked than what would be produced by a pure CC85 outflow. Figure~\ref{fig:xrays} shows that the soft X-rays in the simulation are highly collimated by the disk. This degree of collimation produced in the simulation is likely not realistic, given that the adiabatic nature of the disk leads it to be much puffier at late times than would be expected if it were allowed to cool radiatively.

\section{Convergence}\label{sec:convergence}

The high-resolution simulations in the CGOLS suite are orders of magnitude larger than those that have been used in other contexts to study galactic winds (see Figure~\ref{fig:sim_comparisons}). In this section, we investigate why this level of resolution is necessary. In order to test for convergence of measured quantities, all of the $\Delta x \approx 5\,\mathrm{pc}$ simulations in the CGOLS suite are additionally carried out at both $\Delta x \approx 10\,\mathrm{pc}$ and $\Delta x \approx 20\,\mathrm{pc}$ resolution. We note that these comparison runs are still large simulations - a $\Delta x \approx 20\,\mathrm{pc}$ resolution simulation has a total of $512\times512\times1024 \sim 250$ million cells. Nevertheless, we find that the overall nature of the simulations changes significantly as a result of the $2\times$ improvement in resolution between the A-512 and A-1024 models.

\begin{figure}
\centering
\includegraphics[width=0.75\linewidth]{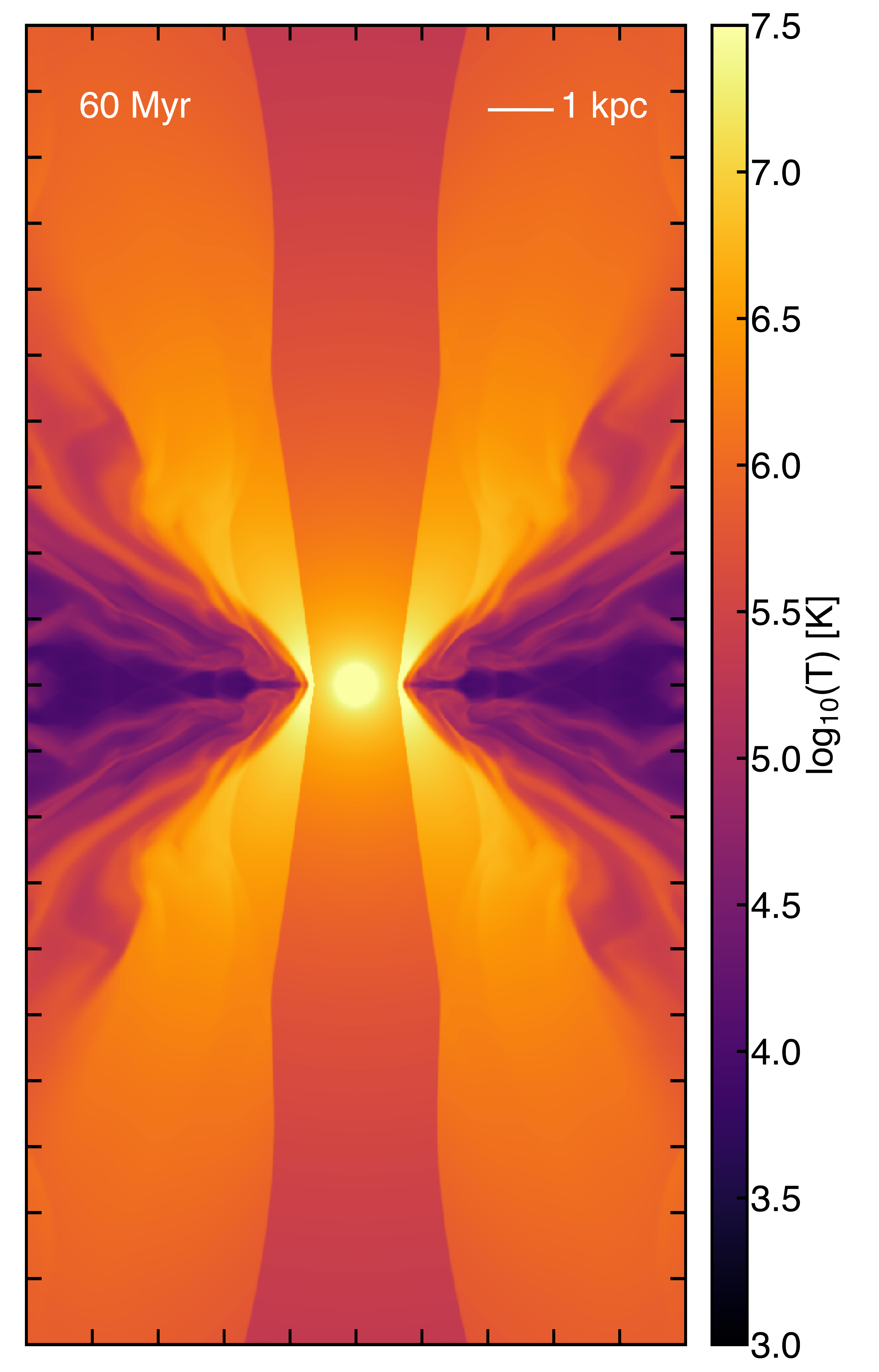}
\caption{Temperature slice at 60 Myr through the $x-z$ plane of the $\Delta x = 20\,\mathrm{pc}$ simulation (model A-512). This slice can be compared directly with Figure~\ref{fig:slices} as an example of the effects of resolution on turbulent features in the disk-outflow interface region.}
\label{fig:Tslice_60_lowres}
\end{figure}

At early times, the results from all three simulations are converged. The radial outflow data presented in Figure~\ref{fig:chevalier_early} for the outflow cone are identical between all three simulations. Other features of the simulations at early times are converged as well, for example, the shock speed of the superbubble as it travels through the ambient CGM and leaves the simulation volume. At later times, we find that the large-scale features remain converged - there is little difference in the radial data plots shown in Figure~\ref{fig:chevalier_late}, for example. However, the degree of turbulence generated in the interface region is quite different for the low-resolution A-512 simulation compared to higher resolutions. As an example, we show in Figure~\ref{fig:Tslice_60_lowres} the 60 Myr temperature slice for the A-512 simulation, which can be directly compared with the slice shown for the fiducial A-2048 simulation in Figure~\ref{fig:slices}. 

\begin{figure}
\centering
\includegraphics[width=1.0\linewidth]{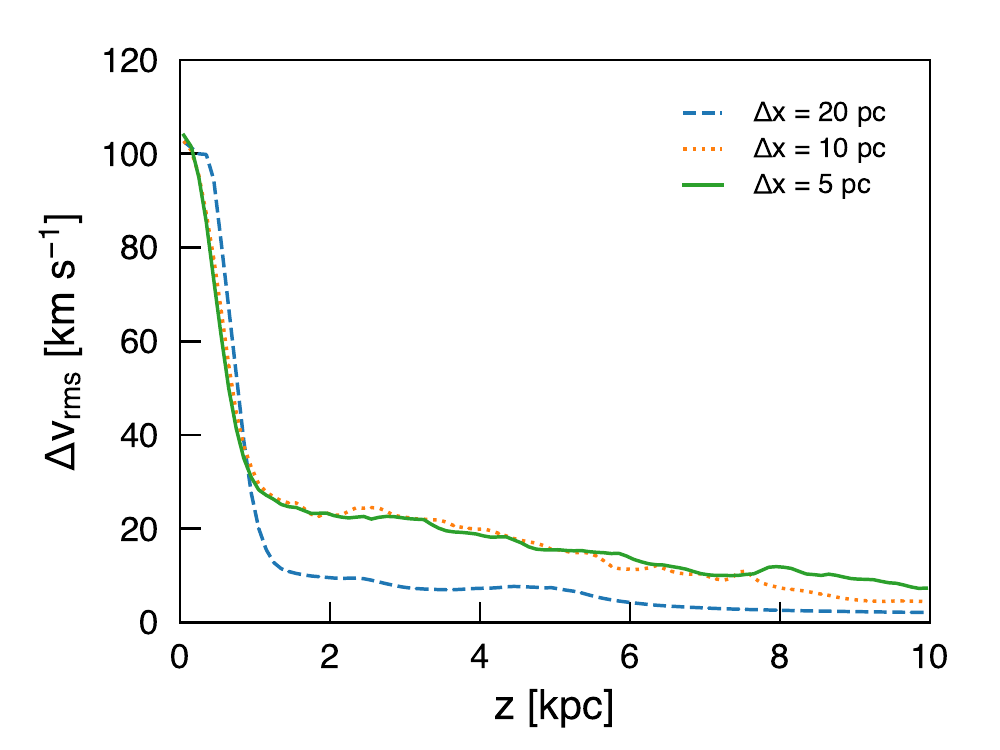}
\caption{Turbulent velocity as a function of height above the disk midplane for simulations with three different resolutions. Turbulent velocity is calculated as the rms average of the difference between the expected radial outflow velocity and the measured velocity at each location.}
\label{fig:deltav_rms}
\end{figure}

We can better quantify the effects of resolution by estimating the degree to which turbulent gas plays a role in each simulation. To do this, we calculate the rms velocity as a function of \added{$z$-}height above the disk midplane. In the nonturbulent case, the velocity at any given point in the outflow is expected to be radial, so to determine the contribution of turbulent gas, we calculate for every cell the magnitude difference between the \added{volume-weighted} total velocity, $v = \sqrt{v_x^2 + v_y^2 + v_z^2}$, and the \added{volume-weighted} radial velocity, $v_r = (v_x x + v_y y + v_z z) / r$. We then determine the rms of this velocity difference,
\begin{equation}
\Delta v_\mathrm{rms} = \sqrt{<(v - v_r)^2>}
\end{equation}
\added{for all cells in the simulation} as a function of \added{$z$-}height above the disk. The resulting profiles are plotted in Figure~\ref{fig:deltav_rms}. At low $z$, $\Delta v_\mathrm{rms}$ primarily shows the average rotation velocity of gas in the disk. Starting at $z \sim 1\,\mathrm{kpc}$, however, the increase in turbulent velocities at higher resolution is clear, and is particularly notable for the highest resolution simulation at large $z$. \added{A mass-weighted version of Figure~\ref{fig:deltav_rms} shows a similar trend, indicating that both the dense gas and the volume-filling gas are turbulent primarily at high resolution.}

\added{Because the outflow velocites are expected to primarily be radial, we interpret $\Delta v_\mathrm{rms}$ as a measure of the turbulence in the outflow. (We have checked that the values of $\Delta v_\mathrm{rms}$ cannot be explained by residual rotational motions of the disk gas.) In these simulations, this turbulence is in part generated by shear instabilities between the radially-moving hot wind and the rotating disk gas. Development of the Kelvin-Helmholtz instability at unresolved interfaces is known to be affected by resolution \citep[e.g.][]{Robertson10}, and the growth rate of the instability is fastest for the smallest wavelength perturbations. As a result, perturbations at the disk-wind interface in the higher resolution simulations will more quickly generate instability on smaller scales. These small eddies are effective in lifting clouds of gas out of the disk and into the outflow, where they contribute to the measured $\Delta v_\mathrm{rms}$. Given the relatively short crossing time for the hot gas to traverse the disk ($\sim 3 - 5\,\mathrm{Myr}$), a resolution of $20\,\mathrm{pc}$ appears insufficient to allow development of the instability at a level that significantly affects turbulent velocities in the outflow.}

\begin{figure}
\centering
\includegraphics[width=1.0\linewidth]{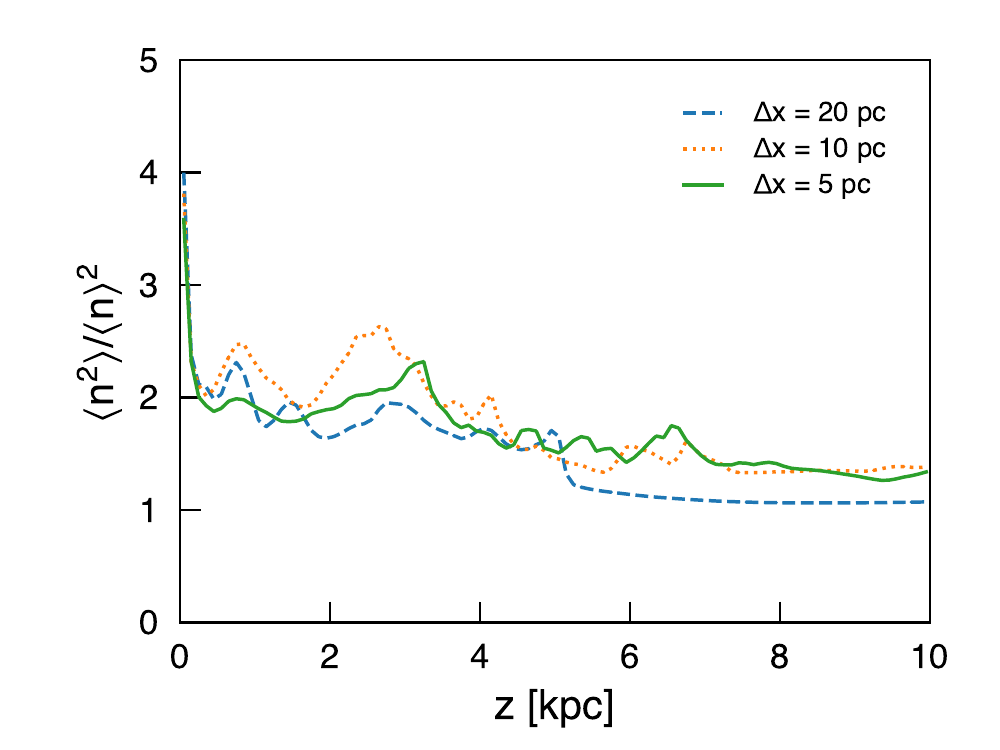}
\caption{Clumping factor of all gas as a function of height above the disk midplane. The three different resolutions have similar values of the clumping factors at all radii. More clumpy structure is observed in the outflow at low $z$.}
\label{fig:clumping}
\end{figure}

\added{In addition to the turbulent velocities, we might also expect the clumpiness of the gas in the outflow to be affected by the resolution of the simulation. In order to check this, we additionally plot in Figure \ref{fig:clumping} the clumping factor of the gas as a function of $z$-height for the three different resolutions. We bin all of the data in the simulations in slices of $\Delta z = 0.1\,\mathrm{kpc}$, and calculate the dimensionless clumping factor for each slice, $\langle n^2 \rangle / \langle n \rangle^2$. Perhaps surprisingly, we see relatively little difference in the clumpiness between the three simulations. As the density slices in Figure \ref{fig:slices} lead us to expect, all three simulations show the most clumping at heights close to the disk, and progressively less as $z$ increases. There may be a slight trend toward increased variability as a function of resolution, and the two higher resolution simulations do show significantly more clumpiness at high latitudes than model A-512.}

\begin{figure*}
\includegraphics[width=0.309\linewidth]{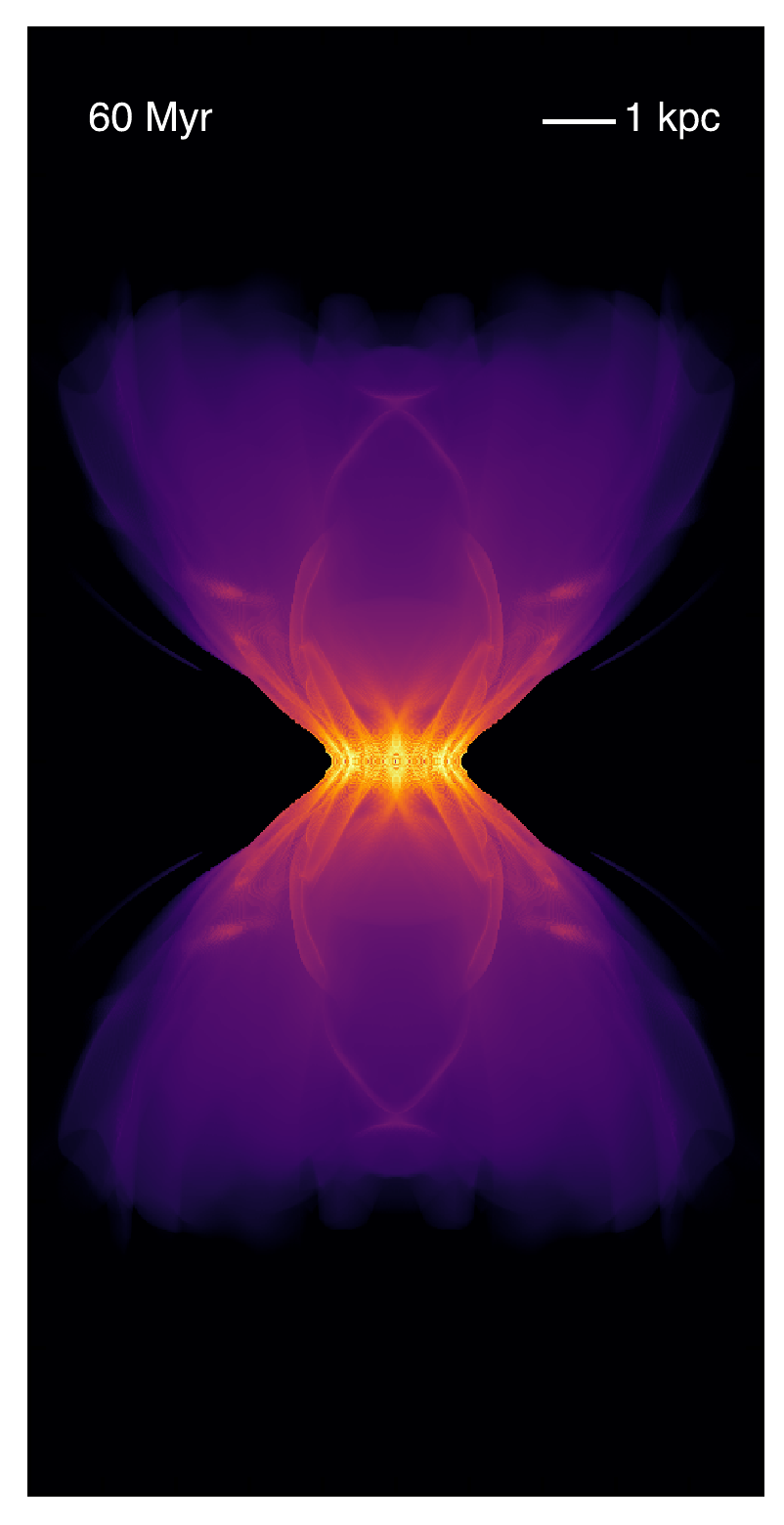}
\includegraphics[width=0.309\linewidth]{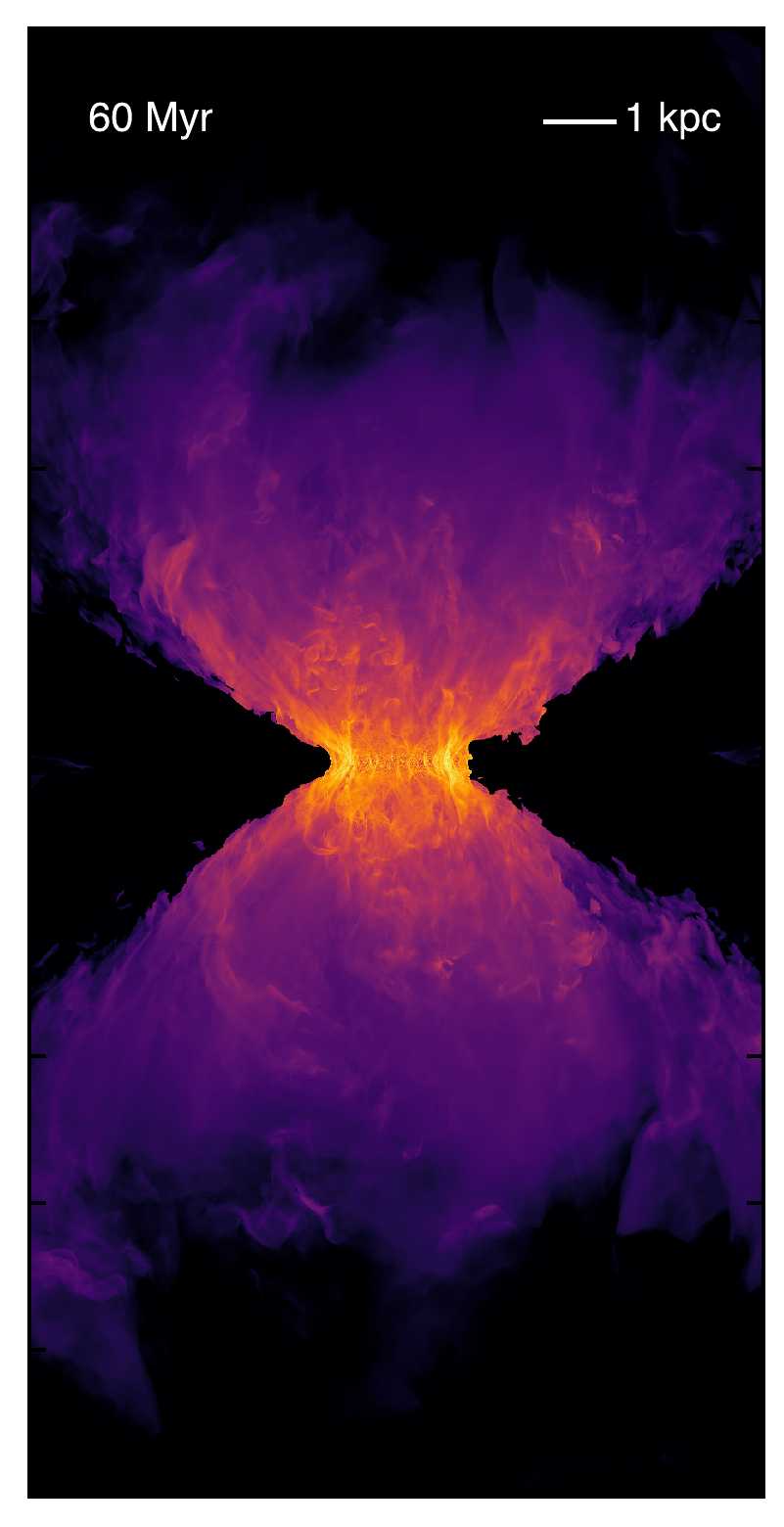}
\includegraphics[width=0.386\linewidth]{2048_xrays_60_xz.pdf}
\caption{Soft X-ray surface brightness maps for simulations at three different resolutions: $\Delta x \approx 20\,\mathrm{pc}$, $\Delta x \approx 10\,\mathrm{pc}$, and $\Delta x \approx 5\,\mathrm{pc}$ (from left to right). The general character of the simulation has not yet converged at $\Delta x = 20\,\mathrm{pc}$ resolution, with large-scale shock features dominating the spatial structure in the X-ray maps, rather than the filamentary structures caused by turbulence in the higher resolution simulations.}
\label{fig:xray_convergence}
\end{figure*}

\added{The turbulence in the outflow is supersonic and produces shocks}. Thus, the increase in turbulent structure as a function of simulation resolution has observable consequences for the spatial distribution of the soft X-ray emitting gas. In Figure~\ref{fig:xray_convergence}, we show the soft X-ray surface brightness maps for the A-512, A-1024, and A-2048 simulations. Clearly, the overall nature of the outflow has not yet converged at $\Delta x = 20\,\mathrm{pc}$ resolution (the left-most map). We note that this is despite the fact that the total size of the simulation volume is much larger than the resolution of a single cell, and the region where the mass and energy is being deposited is also resolved with many cells, given that $R = 300\,\mathrm{pc}$. Large-scale shocks propagating throughout the domain tend to dominate the features seen in the X-ray emission map, and are also the dominant feature in temperature and density projections (not shown).

\added{At the higher resolutions, both the density and temperature projections show a similar filamentary structure, though not as obviously as in the soft X-ray plot. The filaments arise as a result of the shredding and ablation of the denser gas clouds that have been lofted into the outflow from the disk (in movies, we can watch these clouds as they move outward, and more easily observe the expansion and shredding). Many of these clouds  come from the inner region of the disk near the edge of the injection zone, but some additional dense gas is contributed from the shear instabilities at the disk-outflow interface, particularly at later times.}

Whether the total soft X-ray luminosity is converged even in the highest resolution simulation (Model A-2048) is not clear. For the two higher resolution simulations, $L_\mathrm{X}$ tends to increase with resolution. In simulation A-512, the nuclear region and large-scale shock features contribute significantly to the soft X-ray luminosity, such that we find the largest total luminosity in the simulation with the lowest resolution, but the relationship is not monotonic: $L_\mathrm{X} = 2.9\times10^{40}\,\mathrm{erg}\,\mathrm{s}^{-1}$, $L_\mathrm{X} = 1.4\times10^{40}\,\mathrm{erg}\,\mathrm{s}^{-1}$, and $L_\mathrm{X} = 1.9\times10^{40}\,\mathrm{erg}\,\mathrm{s}^{-1}$ for A-512, A-1024, and A-2048, respectively. However, despite the total luminosity in simulation A-512 being the closest to the observed luminosity, the $z$-profile of the emission in the low resolution simulation is much more steeply peaked than is observed. We do not wish to over-interpret these differences in total luminosity, given the uncertainties in the estimated emission caused by our analytic fit to the cooling curve and the simplifying assumption that only cells with volume-average temperature between 0.3 and $2.0\,\mathrm{keV}$ contribute to the X-ray emission. At the very least, we are confident that the match to the \textit{morphology} of the observed soft X-ray brightness profile is improved at higher resolution.

\section{Summary and Conclusions}\label{sec:conclusions}

In this work we have introduced the CGOLS suite, a set of high resolution isolated-galaxy simulations with feedback designed to generate a galactic wind. The volume of the simulation boxes ($10\times10\times20\,\mathrm{kpc}$) allows us to study the evolution of the winds as they escape the galaxy and begin to interact with the CGM, while the resolution ($\Delta x < 5\,\mathrm{pc}$ across the entire volume) is sufficient to capture the small-scale hydrodynamics and thermal instabilities that contribute to mixing and potential formation of multiphase gas.

In addition to a detailed presentation of the features common to all simulations in the suite, we have focused the results of this paper on the first of the CGOLS simulations, an adiabatic simulation of a galaxy with a nuclear starburst modeled after M82 (model A-2048). At early times, the analytic model of a nuclear starburst described by \cite{Chevalier85} fits the data generated by the simulation well, within a biconical region with an opening angle of $\Delta \Omega = 60^\circ$. Outside this region and at late times, significant turbulence is generated in the interface region between the disk gas and the central hot outflow. This region is dominated by gas with densities and temperatures up to an order of magnitude higher than those predicted by the analytic model.

Even with the simplified adiabatic physics of the A-2048 simulation, we calculate a total luminosity in soft X-ray emission that is only a factor of 2 lower than the observed soft X-ray emission in the starburst galaxy M82 (see Figure~\ref{fig:xray_profile}). The turbulent interface region is a key feature in correctly generating the shape of the X-ray profile as a function of height above the disk, given that interactions between the hot, fast wind and slower, denser clouds of gas generate shocks and mixing at large radii that contribute significantly to the X-ray emission.

This paper is the first in a series that will systematically investigate the nature of galactic winds through large-scale global galaxy simulations. Results from additional simulations in the CGOLS suite that include radiative cooling and probe the multiphase nature of the outflow are presented in a companion paper (Schneider et al. \textit{submitted}).

\acknowledgments
We acknowledge inspiration from the Simons Symposia on Galactic Superwinds: Beyond Phenomenology.
This research used resources of the Oak Ridge Leadership Computing Facility, which is a DOE Office of Science User Facility supported under Contract DE-AC05-00OR22725, via an award of computing time on the Titan system provided by the INCITE program (AST125). EES is supported by NASA through Hubble Fellowship grant \#HF-51397.001-A awarded by the Space Telescope Science Institute, which is operated by the Association of Universities for Research in Astronomy, Inc., for NASA, under contract NAS 5-26555. BER is partially supported by NASA grant 17-ATP17-0034, contract NNG16PJ25C, and program HST-GO-14747.

\software{Cholla \citep{Schneider15}; \texttt{matplotlib} \citep{Hunter07}, \texttt{numpy} \citep{VanDerWalt11}, \texttt{hdf5} \citep{hdf5}; Cloudy \citep{Ferland13}}

\begin{appendix}

\section{Modifications to Cholla}\label{app:cholla}

\subsection{Hydrodynamics Solver}

This work used a hydrodynamics integration algorithm not previously incorporated in Cholla, so we document it here for completeness. The solver is based on the description provided in \cite{Stone09} of the ``Van Leer'' integration scheme for the Athena code, a predictor-corrector scheme modeled in turn after the MUSCL-Hancock scheme \citep[see e.g.][]{Toro09}. This second-order integrator updates the vector of conserved variables,
\begin{equation}
\mathbf{u} = [\rho, \rho u, \rho v, \rho w, E]^T
\end{equation}
from time $t = n$ to time $t = n+1$ using fluxes estimated at the half time step $t = n+\frac{1}{2}$. Here $\rho$ is the mass density, $u$, $v$, and $w$ are the three components of the velocity vector, and $E$ is the total energy density. In three dimensions, the update for a cell $(i, j, k)$, can be written:
\begin{equation}
\mathbf{u}^{n+1}_{(i, j, k)} = \mathbf{u}^n + \frac{\Delta t}{\Delta x}\left[\mathbf{F}^{n+\frac{1}{2}}_{(i-\frac{1}{2}, j, k)} - \mathbf{F}^{n+\frac{1}{2}}_{(i+\frac{1}{2}, j, k)}\right] + \frac{\Delta t}{\Delta y}\left[\mathbf{G}^{n+\frac{1}{2}}_{(i, j-\frac{1}{2} k)} - \mathbf{G}^{n+\frac{1}{2}}_{(i, j+\frac{1}{2}, k)}\right] + \frac{\Delta t}{\Delta z}\left[\mathbf{H}^{n+\frac{1}{2}}_{(i, j, k-\frac{1}{2})} - \mathbf{G}^{n+\frac{1}{2}}_{(i, j, k+\frac{1}{2})}\right].
\label{eqn:cell_update}
\end{equation}
Here, $\mathbf{F}^{n+\frac{1}{2}}$, $\mathbf{G}^{n+\frac{1}{2}}$, and $\mathbf{H}^{n+\frac{1}{2}}$ are estimates of the flux between cells at the half time step, $\Delta t$ is the hydrodynamic time step, and $\Delta x$, $\Delta y$, and $\Delta z$ are the cell sizes in each dimension. The basic steps of the three-dimensional algorithm are:
\begin{enumerate}
\item Construct first-order accurate fluxes at each cell interface using the cell-averaged values of the conserved variables and the Riemann solver of choice (in this work we use the HLLC solver). To calculate the first-order flux $\mathbf{F^*}_{i+\frac{1}{2}}$ at the interface between cells $i$ and $i+1$, solve the Riemann problem beginning with left state values $\mathbf{W}_L =(\rho_i, v_i, p_i)$ and right state values $\mathbf{W}_R = (\rho_{i+1}, v_{i+1}, p_{i+1})$.
\item Use these first-order accurate fluxes to update the conserved variables in each cell by a half time step, $\frac{\Delta t}{2}$. Apply the update from Equation~\ref{eqn:cell_update} with $\Delta t = \frac{\Delta t}{2}$ and using $\mathbf{F^*}$, $\mathbf{G^*}$, and $\mathbf{H^*}$.
\item Use the half step values of the conserved variables, $\mathbf{u}^{n+\frac{1}{2}}$ to construct time-centered estimates of the conserved variables at the cell edges, $\mathbf{U}_L^{n+\frac{1}{2}}$ and $\mathbf{U}_R^{n+\frac{1}{2}}$, using the interpolation method of choice (in this work we use a piecewise linear reconstruction with limiting carried out in the conservative variables). Note: unlike PPM \citep{Colella84}, this algorithm does \textit{not} require a time evolution of the reconstructed interface values.
\item Solve a second set of Riemann problems using the time-centered spatially reconstructed variables, $\mathbf{U}_L^{n+\frac{1}{2}}$ and $\mathbf{U}_R^{n+\frac{1}{2}}$, to calculate the time centered second order fluxes, $\mathbf{F}^{n+\frac{1}{2}}$, etc.
\item Use Equation~\ref{eqn:cell_update} with the full time step and the second order fluxes to calculate the new values of the conserved variables at time $n+1$.
\end{enumerate}

In addition to being relatively simple and robust, this algorithm provides a set of first-order fluxes, $\mathbf{F^*}$, $\mathbf{G^*}$, and $\mathbf{H^*}$. When used in Equation~\ref{eqn:cell_update}, these fluxes have the advantage of being positive-definite - that is, in the pure hydrodynamic case, they will not produce negative densities or energies after the cell update. Given the rather extreme nature of the calculations presented in this work, we have found that the full integration scheme can still sometimes produce negative densities or total energies for a given cell when used with higher-order reconstruction methods, particularly in simulations where the gas is also allowed to cool radiatively. In that case, we use the first-order fluxes to perform the complete update, \textit{only} for the affected cell. In practice, this tends to affect of order $1\times10^{-10}$ cells in the calculation on any given update, and was never necessary in the adiabatic simulation presented in this paper.

\subsection{Static Gravity Implementation}

In order to carry out the simulations presented in this work, we added a static gravity module to the Cholla code. The time-averaged gravitational source terms are coupled to the hydrodynamics module using an operator-split cell update, meaning that first each cell is updated using the hydrodynamic fluxes according to Equation~\ref{eqn:cell_update}, then the gravitational source terms are applied to the new momentum and energy. The source terms for momentum and energy are defined as
\begin{equation}
\mathbf{S}_m = \rho \nabla \Phi = \frac{\Delta t}{2}\mathbf{g}(\rho^n + \rho^{n+1}), \quad \mathbf{S}_E = \rho \mathbf{v}\cdot\nabla\Phi = \frac{\Delta t}{4}\mathbf{g}\cdot(\mathbf{v}^n + \mathbf{v}^{n+1})(\rho^n + \rho^{n+1}),
\end{equation}
where $\mathbf{g}$ is the gravitational acceleration calculated from the potential $\Phi$, $\rho^n$ and $\rho^{n+1}$ are the density before and after the hydrodynamic update, and $\mathbf{v}^n$ and $\mathbf{v}^{n+1}$ are the velocity vectors before and after the hydrodynamic update.

\subsection{Cooling Implementation}\label{app:cooling}

\begin{figure}
\centering
\includegraphics[width=0.4\linewidth]{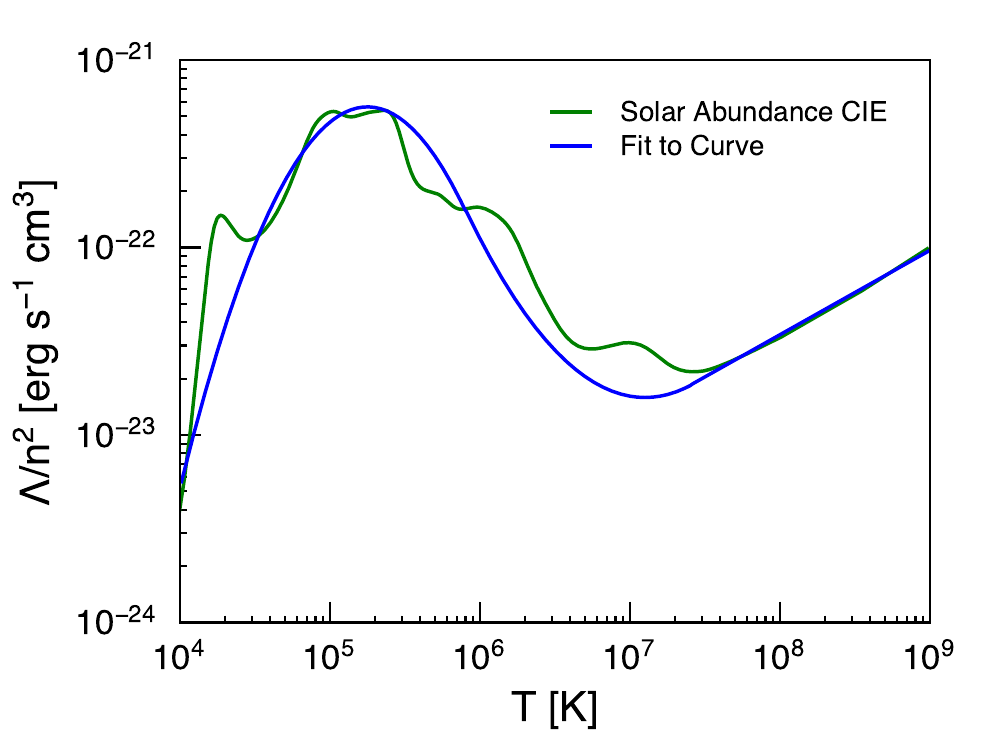}
\caption{Analytic function used to calculate cooling in the CGOLS suite, along with a CIE curve with solar abundances calculated with Cloudy \citep{Ferland13}.}
\label{fig:cooling_curve}
\end{figure}

Although we do not use radiative cooling in the simulation described in this paper, we do use it for the rest of the simulations in the CGOLS suite, so we include our implementation here for completeness. We use an operator-split approach to account for radiative losses in our simulations, as described in Appendix C of \cite{Schneider17}. The only difference is in the calculation of the cooling function, $\Lambda$. Rather than using Cloudy tables as described in that work, here we use a simple analytic function that is a piecewise parabolic fit to a solar metallicity, collisional ionization equilibrium cooling curve computed with Cloudy. The functional form of the cooling curve is
\begin{equation}
\begin{split}
\Lambda &= 0, \quad 10^4 < T \\
\Lambda &= 10^{-1.3(T - 5.25)^2 - 21.25},\quad 10^4 <= T < 10^{5.9} \\
\Lambda &= 10^{0.7(T - 7.1)^2 - 22.8},\quad 10^{5.9} <= T < 10^{7.4} \\
\Lambda &= 10^{0.45T - 26.065},\quad 10^{7.4} <= T,
\end{split}
\end{equation}
where $T$ is in logarithmic units. The cooling curve and fit are shown in Figure~\ref{fig:cooling_curve}. The simulations are run with a temperature floor of $T = 10^4\,\mathrm{K}$, so we do not take into account cooling below this threshold, nor do we account for any radiative heating processes. This cooling curve is also used to calculate the soft X-ray emission maps presented in Section~\ref{sec:xrays}.

\end{appendix}

\bibliography{all}

\end{document}